\documentclass[12pt,journal,onecolumn]{IEEEtran}
\usepackage{epsfig,rotating,setspace,latexsym,amsmath,epsf,amssymb,amsfonts,bm,theorem,epstopdf}
\usepackage{cite,authblk}
\usepackage{algorithmic,algorithm}
\usepackage{setspace}
\usepackage{color}
\usepackage{graphicx}
\usepackage{mathrsfs}

\ifCLASSOPTIONcompsoc
\usepackage[caption=false, font=normalsize, labelfont=sf, textfont=sf]{subfig}
\else
\usepackage[caption=false, font=footnotesize]{subfig}
\fi

\newtheorem{lemma}{Lemma}

\newenvironment{Proof}[1]{\medskip\par\noindent{\bf Proof:\,}\,#1}{{\mbox{\,$\blacksquare$}\par}}

\allowdisplaybreaks

\begin{document}

\title{Minimizing Age of Information with \newline Soft Updates\thanks{This work was supported by NSF Grants CNS 15-26608, CCF 17-13977 and ECCS 18-07348. This paper was presented in part at the Annual Allerton Conference on Communication, Control, and Computing, Monticello, IL, October 2018.}}

\author{Melih Bastopcu \qquad Sennur Ulukus\\
                \normalsize Department of Electrical and Computer Engineering\\
                \normalsize University of Maryland, College Park, MD 20742 \\
                \normalsize {\it bastopcu@umd.edu \qquad {\it ulukus@umd.edu}}}

\maketitle

\vspace*{-1.4cm}

\begin{abstract}
We consider an information updating system where an information provider and an information receiver engage in an update process over time. Different from the existing literature where updates are countable (hard) and take effect either immediately or after a delay, but \emph{instantaneously} in both cases, here updates start taking effect right away but \emph{gradually} over time. We coin this setting \emph{soft updates}. When the updating process starts, the age decreases until the soft update period ends. We constrain the number of times the information provider and the information receiver meet (number of update periods) and the total duration of the update periods. We consider two models for the decrease of age during an update period: In the first model, the rate of decrease of age is proportional to the current age, and in the second model, the rate of decrease of age is constant. The first model results in an exponentially decaying age, and the second model results in a linearly decaying age. In both cases, we determine the optimum updating schemes, by determining the optimum start times and optimum durations of the updates, subject to the constraints on the number of update periods and the total update duration.
\end{abstract}

\section{Introduction}
We consider a system where an information provider updates an information receiver (information consumer) over time. We introduce the concept of \emph{soft updates}, where different from the existing literature where updates are countable (hard) and drop the age instantaneously (possibly after a delay), here, updates are soft and begin reducing the age immediately but drop it gradually over time. Our setting models human interactions where updates are soft, and also social media interactions where an update consists of viewing and digesting many small pieces of information posted, that are of varying importance, relevance and interest to the receiver.

\begin{figure}[t]
	\centerline{\includegraphics[width=0.6\columnwidth]{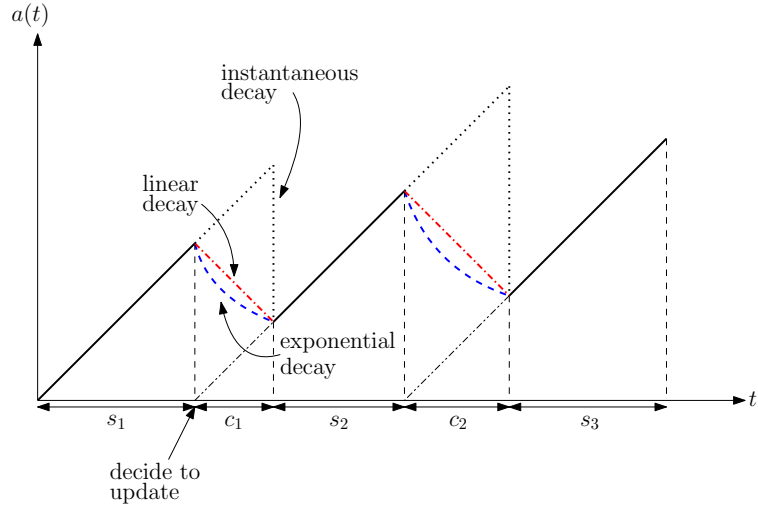}}
	\caption{Update models: Hard updates (instantaneous decay) and soft updates (exponential and linear decay).}
	\label{System_Model}
\end{figure}

Consider a typical information update system as shown in Fig.~\ref{System_Model}. Starting from time zero, information at the receiver gets stale over time, i.e., the age increases linearly. A time comes when the information source decides to update the information receiver. In the existing literature, this is a \emph{hard} update, which is contained in an information packet. This hard update \emph{takes effect} and reduces the age instantaneously to the age of the packet itself at the time of its arrival at the receiver. This is denoted as \emph{instantaneous decay} in Fig.~\ref{System_Model}. The time for the update to take effect (denoted by $c_1$ for the first update) is either random \cite{how_often, multiple_sources, Through_Queues, random_updates, Kam18a, Non_linear, multihop_networks, multi_stream, Update_or_wait, Timely_updates, cache_updating, Wiener_process, Packet_Management}, or fixed and deterministic \cite{Arafa_Age_Energy_Dependent, Arafa_Age}, or zero \cite{Replenishment, Yang_AoI_energy, Arafa_Age_Inc, Arafa_Age, Arafa_Age_Energy_Dependent, Arafa_Age_Online, Arafa18c, Baknina_Age, Baknina_Coded_Upt, Uysal_energy_harvesting, Uysal_finite_Battery, Arafa18d, Yang_noisy_channel, Yang_upt_failure}. Essentially, this is the time for the update packet to \emph{travel} from the transmitter to the receiver, and when it arrives, it drops the age instantaneously. This travel time is random if the update goes through a queue, it is fixed if the update goes through a wireless channel with a non-negligible distance between the transmitter and the receiver, and it is zero if the update goes through a channel with a negligible distance. In contrast, in this work, the soft update begins reducing the age at the time of information source making a decision to update. However, the drop in age is not instantaneous, rather it is \emph{gradual} over time.

We consider two models for the soft update process: In the first model, the rate of decrease in age is proportional to the current age; see (\ref{age-model1}). This is motivated by the fact that new information is most valuable when the current information is most aged, i.e., when the new information is most innovative. This model leads to an exponential decay in the age (denoted by \emph{exponential decay} in Fig.~\ref{System_Model}). Note also that, the exponential decay in the age is consistent with information dissemination in human interactions as well as in social media feeds, where the most important information is conveyed/displayed first, reducing the age faster initially, and less important information is conveyed/displayed next, reducing the age slower subsequently. In the second model, the rate of decrease in age is not a function of the current age, rather it is constant; see (\ref{age-model2}). In this case, the age decreases linearly (denoted by \emph{linear decay} in Fig.~\ref{System_Model}).

In this paper, we determine the optimum updating schemes for \emph{soft update systems}. We are given the total system duration over which the average age is calculated $T$, the number of update periods (i.e., the number of times information provider and information receiver are allowed to meet) $N$, and the total allowed update duration $T_c$. We solve for the optimum start times of the soft updates and their optimum durations in order to minimize the overall age.

We show that for both exponentially and linearly decaying age models, the optimal policy is to have exactly $N$ soft updates, completely utilize the given total update duration $T_c$, and divide the total update duration $T_c$ equally among $N$ updates. We note that when $T_c$ is large compared to $T$, we may have multiple optimal solutions. In order to generalize the solution for both models and for any $T_c$, we choose the optimal policy which allocates equal amount of time for each soft update. For the exponentially decaying age model, if $T_c$ is small compared to $T$, the optimal policy schedules the updates regularly; if $T_c$ is large enough, the system starts updating at time zero, proceeds to update continually until $T_c$ is completely utilized, and lets age grow then on until the end. For the linearly decaying age model, if $T_c$ is small compared to $T,$ the optimal policy schedules the updates regularly and the age after each soft update goes down exactly to zero; if $T_c$ is large enough, age not only goes down to zero after each soft update, but also stays at zero for some time after each soft update. In addition, for the exponentially decaying age model with small $T_c$ and for the linearly decaying age model for all $T_c$, we show that the resulting age decreases with $N$.

Finally, we provide numerical results where not only the number of soft update opportunities and the total duration of soft updates are constrained, but also the time periods during which update encounters may take place are constrained as well.

\section{System Model and the Problem}
Let $a(t)$ be the instantaneous age at time $t$. Without loss of generality, let $a(0) = 0$. When there is no update, the age increases linearly with time. We consider two different soft update models. In the first model, the rate of decrease in age is proportional to the current age:
\begin{align} \label{age-model1}
\frac{da(t)}{dt} = -\alpha a(t)
\end{align}
where $\alpha$ is a fixed constant. In this model, the age decreases exponentially during a soft update period. In the second model, the rate of decrease in age does not depend on the current age, instead it remains constant:
\begin{align} \label{age-model2}
\frac{da(t)}{dt} = -\alpha
\end{align}
where $\alpha$ is a fixed constant. In this model, the age decreases linearly during a soft update period.

Let us denote the beginning of the $i$th soft update period by $t_i$ and the end of the $i$th soft update period by $t_i'$. Then, the age evolves as:
\begin{align}
a(t)& \triangleq
\begin{cases}
a(t_{i-1}')+t-t_{i-1}',&  t_{i-1}'<t < t_{i}\\
f (a(t_i),\alpha,t),& t_i < t<t_i'
\end{cases}
\end{align}
where $f(a(t_i),\alpha,t) = a(t_i)e^{-\alpha (t-t_i)}$ for the exponentially decaying age model, and $f(a(t_i),\alpha,t) = (a(t_i)-\alpha (t-t_i))^+$ for the linearly decaying age model, where $(x)^+ = x$ for $x>0$ and $(x)^+ = 0$ for $x\leq 0$. For both models, if the current age is larger than zero, age decreases during an update period. For the linearly decaying age model, depending on the update duration and the age at the beginning of the update, the age can go down to zero. Here, we consider the most general case where the age can stay at zero if the duration of the update period is large enough.\footnote{In \cite{soft_upt_allerton}, for the linearly decaying age model, we consider the case where we terminate an update process if the current age goes down to zero. In this paper, we assume that an update process can continue after the current age becomes zero. During this period, since the update process is on, the system does not age, i.e., the age stays at zero.} For the exponentially decaying age model, age stays at zero only if we have an update starting at time $t=0$. Otherwise, age never goes down to zero in a finite update duration.

Our objective is to minimize the average age of information (AoI) of the system subject to a total of $N$ soft update periods, a total update duration of $T_c$, over a total session duration of $T$. We formulate the problem as:
\begin{align}
\label{problem_1}
\min_{\{t_{i}, t_{i}' \}}  \quad & \frac{1}{T} \int_{0}^{T} a(t) dt \nonumber \\
\mbox{s.t.} \quad & \sum_{i=1}^{N} (t_{i}'-t_{i}) \leq T_c
\end{align}

\begin{figure}[t]
	\centerline{\includegraphics[width=0.6\columnwidth]{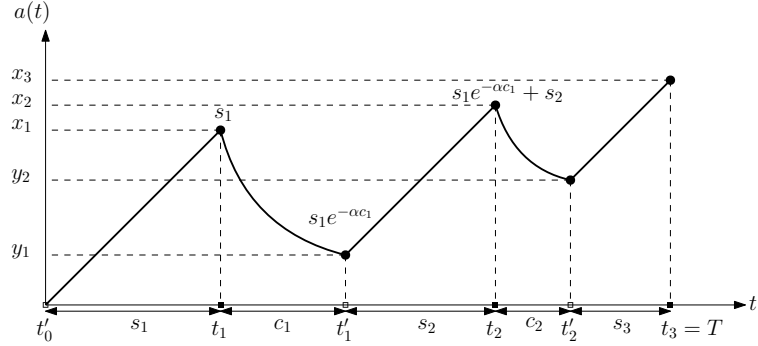}}
	\caption{A general example evolution of age in the case of exponentially decaying age.}
	\label{gradual}
\end{figure}

We define the duration of the $i$th update period as $c_i= t_i'-t_i$, and the $i$th aging period as $s_i= t_{i}-t_{i-1}'$. For convention, we let $t_0' =0$, and $t_{N+1}=T$. Additionally, we denote the age at the beginning of the $i$th soft update period by $x_i$, and the age at the end of the $i$th soft update period by $y_i$. Therefore, we obtain three equivalent sets of variables to describe the system: $\{t_i,t_i'\}_{i=1}^{N}$, $\{s_i,c_i\}_{i=1}^{N}$, and $\{x_i,y_i\}_{i=1}^{N}$. We retain these three sets of equivalent variables throughout the paper; we find it more convenient to express $A_T$ in terms of $x_i$ and $y_i$ for the exponentially decreasing age model, and in terms of $s_i$ and $c_i$ for the linearly decreasing age model. The relationship between $(t_i,t_i')$, $ (s_i,c_i)$, and $(x_i,y_i)$ is shown in Fig.~\ref{gradual}.

Let $A_T \triangleq \int_{0}^{T} a(t) dt$ be the total age. Note that minimizing $\frac{A_T}{T}$ is equivalent to minimizing $A_T$ since $T$ is a known constant. In the following sections, we provide the optimal policies that minimize the age for the cases of exponentially and linearly decaying age models.

\section{Exponentially Decaying Age Model} \label{sect:gradual}
In the case of exponentially decaying age, the age function evolves as in Fig.~\ref{gradual}. Age, in this
case, is given in terms of $x_i$ and $y_i$ as:
\begin{align}
A_T = \sum_{i=1}^{N} \frac{x_i^2}{2}-\frac{y_i^2}{2}+\frac{1}{\alpha}(x_i-y_i)+\frac{x_{N+1}^2}{2}\label{A_T_exp}
\end{align}
We minimize $A_T$ in (\ref{A_T_exp}) by choosing $x_i$ and $y_i$, equivalently, by choosing $t_i$ and $t_i'$, and $s_i$ and $c_i$, for all $i$. In the following lemma, we show that the total update time, $T_c$, should be fully utilized.
\begin{lemma} \label{lemma1}
For the exponentially decaying age model, in the optimal policy, we must have $\sum_{i=1}^{N}c_i = T_c$.
\end{lemma}
\begin{Proof}
We prove this by contradiction. Assume that there exists an optimal policy such that $\sum_{i=1}^{N}c_i < T_c$. Then, we can simply obtain another feasible policy by increasing one of the $c_i$
and decreasing one of the $s_j$. Note that this new policy yields a smaller age. Thus, we reached
a contradiction, and $\sum_{i=1}^{N}c_i = T_c$ must be satisfied.
\end{Proof}

Thus, from Lemma \ref{lemma1}, the total update time should be fully used. Then, we need
to determine when to start a soft update and the duration of each soft update. In the case of
$T_c = T$, the optimal policy is to start updating at $t = 0$ and continue to update until $t = T$.
The optimal age in this case is $A_T = 0$. Thus, for the rest of this section, we consider the case
where $T_c < T$. We formulate the optimization problem as:
\begin{align}
\label{problem_1_exp}
\min_{\{x_{i}, y_{i} \}}  \quad & \sum_{i=1}^{N} \frac{x_i^2}{2}-\frac{y_i^2}{2}+\frac{1}{\alpha}(x_i-y_i)+\frac{x_{N+1}^2}{2} \nonumber \\
\mbox{s.t.} \quad & \sum_{i=1}^{N} \frac{1}{\alpha}\log\left(\frac{x_i}{y_i}\right) \leq T_c \nonumber \\
\quad & \sum_{i=1}^{N} \left(x_i-y_i+\frac{1}{\alpha}\log\left(\frac{x_i}{y_i}\right)\right)+x_{N+1} = T \nonumber \\
\quad & y_i \leq x_i, \quad y_i \leq x_{i+1}, \quad x_i\geq 0, \quad y_i \geq 0
\end{align}
where the cost function is the age expression in (\ref{A_T_exp}); the first constraint is the constraint on the total soft update duration which is obtained by noting that the $i$th update duration $c_i$ is expressed in terms of $x_i$ and $y_i$ as $y_i = x_i e^{-\alpha c_i}$ therefore, $c_i=\frac{1}{\alpha}\log\left(\frac{x_i}{y_i}\right)$; the second constraint is the total session duration constraint which is the sum of aging durations $s_i$ and update durations $c_i$ where $s_i$ is given in terms of $x_i$ and $y_i$ as $s_i = x_i-y_{i-1}$ with the convention of $y_0=0$; and the third (last) set of constraints state that age in the update period decreases ($y_i \leq x_i$), age in the aging period increases ($y_i\leq x_{i+1}$), and age at all times is non-negative ($x_i\geq 0$, $y_i\geq 0$).

We write the Lagrangian for the problem in (\ref{problem_1_exp}) as:
\begin{align}
\mathcal{L} =& \sum_{i=1}^{N} \frac{x_i^2}{2}-\frac{y_i^2}{2}+\frac{1}{\alpha}(x_i-y_i)+\frac{x_{N+1}^2}{2}+\lambda\left(\sum_{i=1}^{N} \frac{1}{\alpha}\log\left(\frac{x_i}{y_i}\right)- T_c\right) \nonumber\\
&+\beta \left(T-\left(\sum_{i=1}^{N} \left(x_i-y_i +\frac{1}{\alpha}\log\left(\frac{x_i}{y_i}\right)\right) +x_{N+1}\right) \right) \nonumber\\
&+\sum_{i=1}^{N}\gamma_i(y_i-x_i) +\sum_{i=1}^{N}\theta_i(y_i-x_{i+1})-\sum_{i=1}^{N+1} \mu_ix_i-\sum_{i=1}^{N}\nu_iy_i
\end{align}
where $\lambda\geq0$, $\gamma_i\geq0$, $\theta_i\geq0$, $\mu_i\geq 0$, $\nu_i \geq 0$, and $\beta$ can be anything.
Note that the problem given in (\ref{problem_1_exp}) is not convex. Thus, KKT conditions are necessary but not sufficient for the optimal solution. The KKT conditions are:
\begin{align}
\frac{\partial \mathcal{L}}{\partial x_1} = & x_1+\frac{1}{\alpha}+\frac{\lambda}{\alpha x_1}-\beta\left(1+\frac{1}{\alpha x_1} \right)-\gamma_1-\mu_1 =0 \label{part_x1}\\
\frac{\partial \mathcal{L}}{\partial x_i} = & x_i+\frac{1}{\alpha}+\frac{\lambda}{\alpha x_i}-\beta\left(1+\frac{1}{\alpha x_i} \right)-\gamma_i-\theta_{i-1}-\mu_i =0, \quad i = 2,\dots, N \label{part_x2}\\
\frac{\partial \mathcal{L}}{\partial x_{N+1}} = & x_{N+1}-\beta-\theta_N-\mu_{N+1} =0 \label{part_x4}\\
\frac{\partial \mathcal{L}}{\partial y_i} = & -y_i-\frac{1}{\alpha}-\frac{\lambda}{\alpha y_i}+\beta\left(1+\frac{1}{\alpha y_i} \right)+\gamma_i+\theta_i-\nu_i =0, \quad i = 1,\dots,N \label{part_c1}
\end{align}
The complementary slackness conditions are:
\begin{align}
\lambda\left(\sum_{i=1}^{N} \frac{1}{\alpha}\log\left(\frac{x_i}{y_i}\right)- T_c\right) &= 0 \label{eqn_lamd}\\
\beta \left(T-\left(\sum_{i=1}^{N} \left(x_i-y_i+\frac{1}{\alpha}\log\left(\frac{x_i}{y_i}\right)\right)+x_{N+1}\right) \right) & = 0\label{eqn_beta}\\
\gamma_i(y_i-x_i) & =0\\
\theta_i(y_i-x_{i+1}) & =0\\
\mu_i x_i & = 0\\
\nu_i y_i & = 0
\end{align}
In the following, we consider two cases separately: $x_1>0$ and $x_1=0$ in the optimal solution. First, we investigate the case when $x_1>0$.

\subsection{The Optimal Solution Structure When $x_1>0$} \label{exponential-x1-positive}
Since $x_1>0$, from the complementary slackness conditions, we have $\mu_1 = 0$. Since $y_1=x_1e^{-\alpha c_1}$, we have $y_1>0$. Due to $x_2\geq y_1$, we have $x_2>0$. Continuing similarly, we have $y_i>0$ and $x_i>0$ for all $i$. Thus, $\mu_i = 0$ and $\nu_i = 0$ for all $i$. In addition, due to Lemma~\ref{lemma1}, there exists at least one $i$ such that $x_i>y_i$. For these cases, $\gamma_i = 0$. Since $T>T_c$, we have at least one $j$ such that $x_{j+1}>y_j$ and corresponding $\theta_j = 0$. Then, we have four possible cases. Next, we investigate them separately.

\subsubsection{Case A: $x_i>y_i$ and $x_{i+1}> y_i$ for all $i$}\label{CaseA}
\begin{figure}
 	\subfloat[\label{case_a}]{%
 		\includegraphics[width=0.45\linewidth]{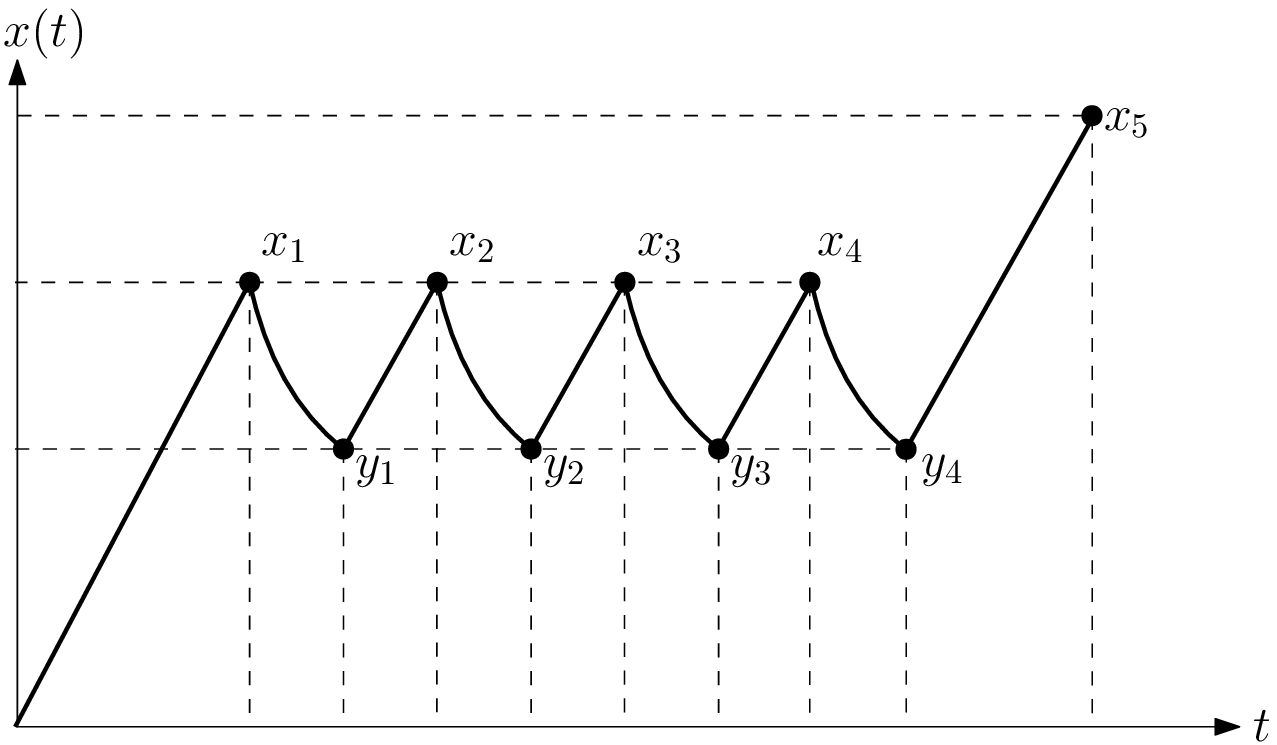}}
 	\hfill
 	\subfloat[\label{x_i_y_i}]{%
 		\includegraphics[width=0.45\linewidth]{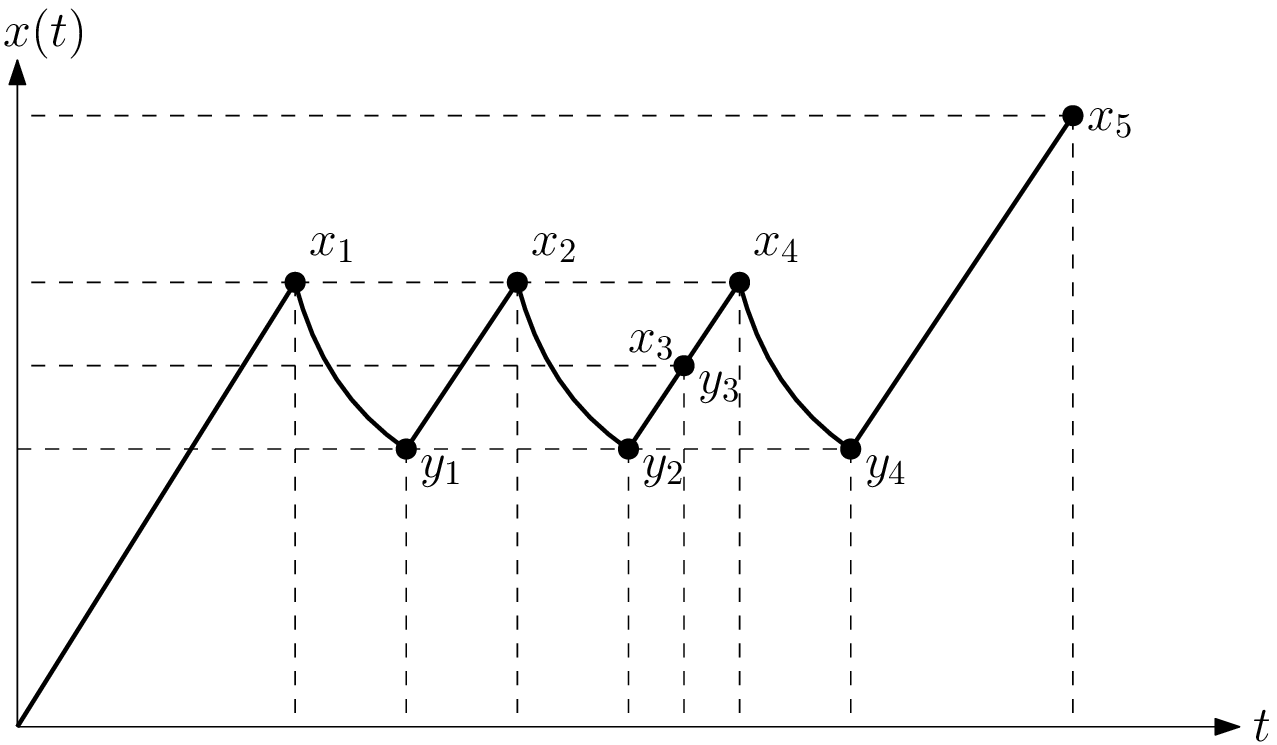}}\\
 	\subfloat[\label{x_i_1_c_i}]{%
 		\includegraphics[width=0.45\linewidth]{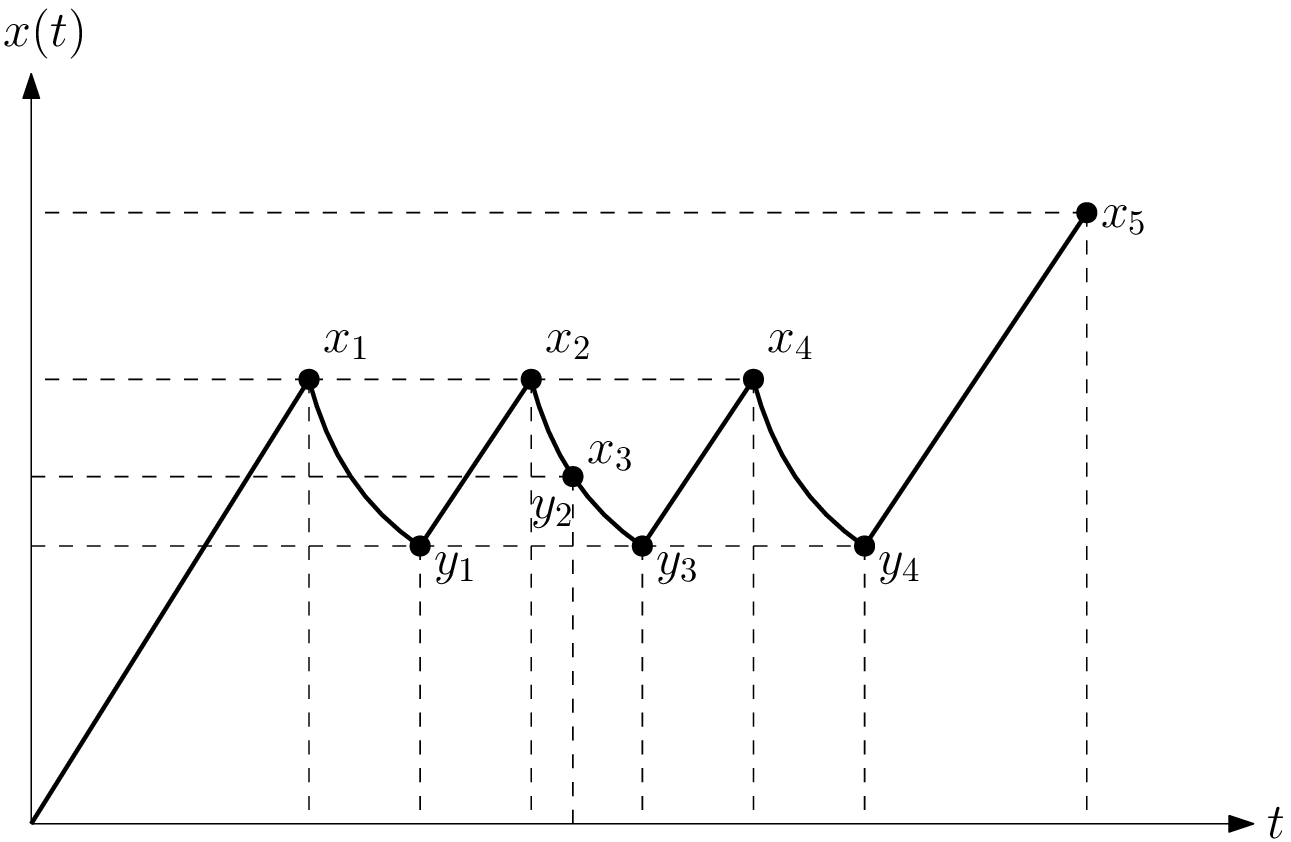}}
 	\hfill
 	\subfloat[\label{both_cases}]{%
 		\includegraphics[width=0.45\linewidth]{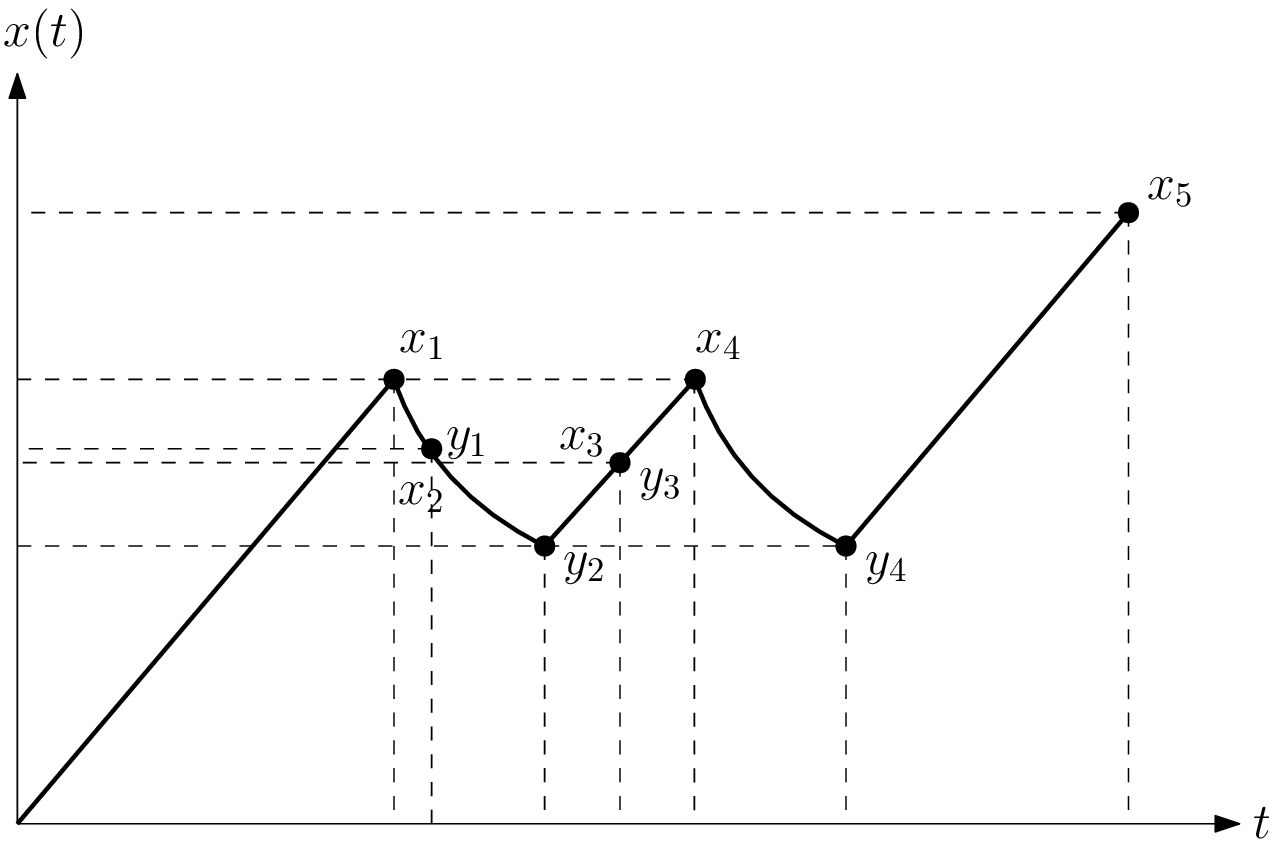}}
 	\caption{Depiction of the cases for the exponentially decaying age model with $x_1>0$ (a) where $ x_i> y_i$ and $x_{j+1} >y_j$ for all $i$ and $j$, (b) where $x_i =y_i$ for some $i$, (c) where $x_{j+1} =y_j$ for some $j$, (d) where $x_i =y_i$ and $x_{j+1} =y_j$ for some $i$ and $j$. }
 	\label{allcases}
\end{figure}

In this case, we have $N$ strict updating and correspondingly $N+1$ strict aging periods. This case is shown in Fig.~\ref{allcases}(a). Since $x_i>y_i$ and $x_{i+1}>y_i$ for all $i$, from the complementary slackness conditions, we have $\gamma_i =0$ and $\theta_i = 0$ for all $i$. Thus, (\ref{part_x1})-(\ref{part_c1}) become:
\begin{align}
\frac{\partial \mathcal{L}}{\partial x_i} = & x_i+\frac{1}{\alpha}+\frac{\lambda}{\alpha x_i}-\beta\left(1+\frac{1}{\alpha x_i} \right) =0, \quad i =1,\dots,N \label{part_x1_s}\\
\frac{\partial \mathcal{L}}{\partial x_{N+1}} = & x_{N+1}-\beta=0\label{part_x4_s}\\
\frac{\partial \mathcal{L}}{\partial y_i} = & -y_i-\frac{1}{\alpha}-\frac{\lambda}{\alpha y_i}+\beta\left(1+\frac{1}{\alpha y_i} \right) =0 , \quad i =1,\dots,N \label{part_c1_s}
\end{align}
Note that the right hand sides of $\frac{\partial \mathcal{L}}{\partial x_i}$ and $\frac{\partial \mathcal{L}}{\partial y_i}$ in (\ref{part_x1_s}) and (\ref{part_c1_s}) are the same second degree equalities. Since we consider the case where $x_i>y_i$ for all $i$, the larger root of this equality gives $x_i$ and the smaller root gives $y_i$. Rewriting (\ref{part_x1_s}) in terms of a single variable $z$, we have,
\begin{align}
z+\frac{1}{\alpha}+\frac{\lambda}{\alpha z}-\beta\left(1+\frac{1}{\alpha z} \right) =0
\end{align}
which is equivalent to,
\begin{align}
\alpha z^2+z(1-\beta\alpha)+(\lambda-\beta) = 0 \label{eqn_with_z}
\end{align}
The roots of this equation are,
\begin{align}
z_1 = \frac{-(1-\beta\alpha)+\sqrt{(1-\alpha\beta)^2-4\alpha(\lambda-\beta)}}{2\alpha}\\
z_2 = \frac{-(1-\beta\alpha)-\sqrt{(1-\alpha\beta)^2-4\alpha(\lambda-\beta)}}{2\alpha}
\end{align}
and we have $x_i  = z_1$ and $y_i = z_2$, for all $i$. Note that in order to have two positive roots, we need $1-\beta \alpha <0$. Thus, we have:
\begin{align}
x_{N+1} = \beta > \frac{1}{\alpha}\label{x_4_eqn1}
\end{align}
where we also used (\ref{part_x4_s}). Since $c_i = \frac{1}{\alpha}\log\left(\frac{x_i}{y_i}\right)$ and $\sum_{i=1}^{N}c_i = T_c$ and since all $x_i$ are equal among themselves and all $y_i$ are equal among themselves, we have all $c_i$ equal and $c_i = \frac{T_c}{N}$.

Next, we note from (\ref{eqn_with_z}) that,
\begin{align}
x_{N+1} = x_i+y_i+\frac{1}{\alpha} \label{x_N+1}
\end{align}
Further, by using (\ref{eqn_lamd}), (\ref{eqn_beta}) and Lemma~\ref{lemma1}, we obtain,
\begin{align}
N(x_i-y_i)+x_{N+1} =T-T_c\label{x_4_eqn2}
\end{align}
Substituting (\ref{x_N+1}) into (\ref{x_4_eqn2}), and noting that $\frac{1}{\alpha}\log \left(\frac{x_i}{y_i}\right) =c_i = \frac{T_c}{N}$, we solve for $x_i$ as,
\begin{align}
x_i = \frac{T-T_c-\frac{1}{\alpha}}{(N+1)-(N-1)e^{-\frac{\alpha T_c}{N}}}, \quad i=1,\dots,N \label{x1_closed}
\end{align}
and
\begin{align}
x_{N+1} = \frac{(T-T_c)\left(1+e^{-\frac{\alpha T_c}{N}}\right)+\frac{N}{\alpha}}{(N+1)-(N-1)e^{-\frac{\alpha T_c}{N}}}
\end{align}

\begin{figure}[t]
	\centerline{\includegraphics[width=0.6\columnwidth]{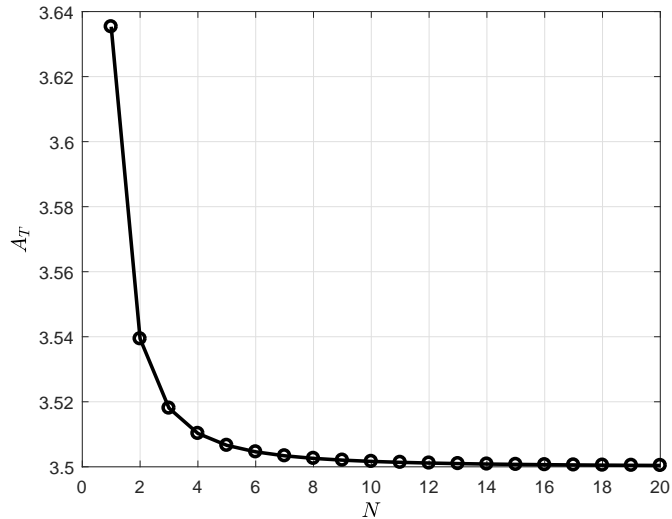}}
	\caption{Minimum age as a function of $N$ in the exponentially decaying age case for $T=5$, $T_c=2$, and $\alpha=1$.}
	\label{exp_case}
\end{figure}

With this solution, the minimum age, $A_T$, is:
\begin{align}
A_T = \frac{1}{2}\left(T-T_c-\frac{1}{\alpha}\right)^2\frac{1+e^{-\frac{\alpha T_c}{N}} }{(N+1)-(N-1)e^{-\frac{\alpha T_c}{N}}}+\frac{1}{\alpha}(T-T_c)-\frac{1}{2\alpha^2} \label{A_t_closed}
\end{align}
We note that $A_T$ is monotonically decreasing with respect to $N$ in \textit{Case A}. To see this, we note that the derivative of $A_T$ in (\ref{A_t_closed}) with respect to $N$ is equal to,
\begin{align}
\frac{\partial A_T}{\partial N} = C\frac{2\left(\frac{\alpha T_c}{N}\right) e^{-\frac{\alpha T_c}{N}}+e^{-\frac{2\alpha T_c}{N}}-1}{\left((N+1)-(N-1)e^{-\frac{\alpha T_c}{N}}\right)^2} \label{A_t_der}
\end{align}
where $C= \frac{1}{2}\left(T-T_c-\frac{1}{\alpha}\right)^2$. Note that $C$ and the denominator in (\ref{A_t_der}) are always positive.
Next, letting $a= \frac{\alpha T_c}{N}$, the numerator of (\ref{A_t_der}) becomes $ 2ae^{-a}\left(1-\frac{\sinh(a)}{a}\right)$. Since $\sinh(a)\geq a$ for all $a\geq0$, and therefore, $\frac{\sinh(a)}{a}\geq 1$ for all $a\geq0$, this implies that the numerator of (\ref{A_t_der}) is always negative, implying that $\frac{dA_T}{dN}\leq 0.$ As an aside, we plot $A_T$ versus $N$ for $T=5$, $T_c =2$, and $\alpha = 1$ in Fig.~\ref{exp_case}. Note that $A_T$ is a decreasing function with respect to $N$ with a limit:
\begin{align}
\lim\limits_{N\rightarrow \infty} A_T =  \frac{1}{2}\left(T-T_c-\frac{1}{\alpha}\right)^2\frac{2 }{2+\alpha T_c}+\frac{1}{\alpha}(T-T_c)-\frac{1}{2\alpha^2}
\end{align}

\subsubsection{Case B: $x_i=y_i$ for some $i$ and $x_{j+1}> y_j$ for all $j$}
This case is shown in Fig.~\ref{allcases}(b). This is equivalent to \textit{Case A} with $N' = N-n$, where $n$ is the total number of update processes with $x_i=y_i$. We know from \textit{Case A} that $A_T$ decreases with $N$. Thus, \textit{Case B} cannot be optimal.

\subsubsection{Case C: $x_i>y_i$ for all $i$ and $x_{j+1}= y_j$ for some $j$}
This case is shown in Fig.~\ref{allcases}(c). Similar to \textit{Case B}, this case is equivalent to \textit{Case A} with $N' = N-m$, where $m$ is the total number of aging processes with $x_{j+1}=y_j$. Thus, \textit{Case C} cannot be optimal.

\subsubsection{Case D: $x_i=y_i$ for some $i$ and $x_{j+1}= y_j$ for some $j$}
This case is shown in Fig.~\ref{allcases}(d). This is equivalent to \textit{Case A} with $N'= N-k$, where $k$ is the total number of update and aging processes with $x_i=y_i$ and $x_{j+1}= y_j$ subtracting $i=j$ cases. Thus, \textit{Case D} cannot be optimal.

Thus, we see that if we have $x_1>0$, the optimal solution only comes from \textit{Case A}. In addition, from (\ref{x_4_eqn1}) and (\ref{x_4_eqn2}), in order to have $x_1>0$, we need:
\begin{align}
\frac{1}{\alpha} < x_{N+1} < T-T_c
\end{align}
Therefore, in the optimal solution, if we have $x_1>0$, then $T>T_c+\frac{1}{\alpha}$ should be satisfied.

As a result, if $x_1>0$ in the optimal solution, this should happen for $T$ and $T_c$ that satisfy $T>T_c+\frac{1}{\alpha}$, i.e., $T_c$ is relatively small in relation to $T$, and in this case, the optimal solution is to update $N$ times with equal update durations, i.e., $c_i=\frac{T_c}{N}$, for all $i$ as shown in Fig.~\ref{allcases}(a).

Next, we study the optimal solution structure when $x_1=0$.

\subsection{The Optimal Solution Structure When $x_1 =0$} \label{exponential-x1-zero}
So far, we studied the optimal solution structure when $x_1>0$. We see that this case requires $T>T_c+\frac{1}{\alpha}$. Thus, when $T\leq T_c+\frac{1}{\alpha}$, we have $x_1 = 0$. Since $y_1 = x_1 e^{-\alpha c_1}$, we have $y_1=0$. In the following, we show that if $T\leq T_c+\frac{1}{\alpha}$, then the optimal policy is to keep the age equal to zero starting from $t=0$ till $t=T_c$, and let the age grow from $t=T_c$ till $t=T$.

We see that if $T\leq T_c+\frac{1}{\alpha}$, then $x_1= y_1 = 0$. Also, $0\leq c_1 \leq T_c$. If $c_1=T_c$, then the optimal policy is exactly as descibed above, i.e., start the update policy at $t=0$ and continue updating until $t=c_1=T_c$, and stop updating then, i.e., let the age grow until $t=T$. If $c_1<T_c$, we need to first show that $x_2= 0$, and therefore, $y_2= 0$. We prove this by contradiction. Assume that there exists an optimal policy such that $T \leq T_c+\frac{1}{\alpha}$, $x_1 =y_1 = 0$, and $x_2>0$. Since the age stays at zero during $c_1$, we can formulate a new age minimization problem starting from $t= c_1$. For the new problem, $T' = T-c_1$, $T_c' = T_c-c_1$, and $N'=N-1$. Since $T' =T- c_1 \leq T_c-c_1+\frac{1}{\alpha}=T_c'+\frac{1}{\alpha}$, we have $T'\leq T_c'+\frac{1}{\alpha}$. Thus, for the new problem, we reach a contradiction and we must have $x_2 = 0$ as well as $y_2=0$. At this point, we have $0\leq c_2 \leq T_c-c_1$. If $c_2= T_c-c_1$, we have the desired policy described above. If not, we repeat the same steps to argue that $x_3=0$, and thus, $y_3 =0$. Then, we select $c_3\in[0,T_c-c_1-c_2]$. Thus, for the remaining terms, we can either argue that $c_i=T_c-\sum_{j=1}^{i-1}c_j$ or show that $x_{i+1}=y_{i+1} =0$ and select $c_{i+1}$ accordingly. At the end, the optimal policy is to update starting from $t=0$, proceed to update continually until $t=T_c$, and then let the age grow until $T$.

Here, we may view the optimal solution in multiple ways: We may view it as a single update that lasts $c_1=T_c$ second, or we may view it $N$ updates that altogether last $c_1+\dots+c_N=T_c$ seconds, or $N'$ updates where $1<N'<N$ with appropriate selection of corresponding $c_i$ to sum up to $T_c$. Even though we have such multiple optimal solutions, we choose the one with $N$ updates with equal update durations (to be consistent with the solution in the previous sub-section), i.e., $c_i=\frac{T_c}{N}$, for all $i$. Thus, we have $x_i=y_i=0$ for $i=1,\dots,N$ and $x_{N+1} = T-T_c$.

With this solution, the minimum age, $A_T$, is:
\begin{align}
A_T = \frac{(T-T_c)^2}{2} \label{A_t_closed_1}
\end{align}
We note that $A_T$ in (\ref{A_t_closed_1}) does not decrease with $N$ unlike $A_T$ in (\ref{A_t_closed}).

\begin{figure}
	\centering
	\subfloat[\label{opt_soln_exp}]{%
		\includegraphics[width=0.43\linewidth]{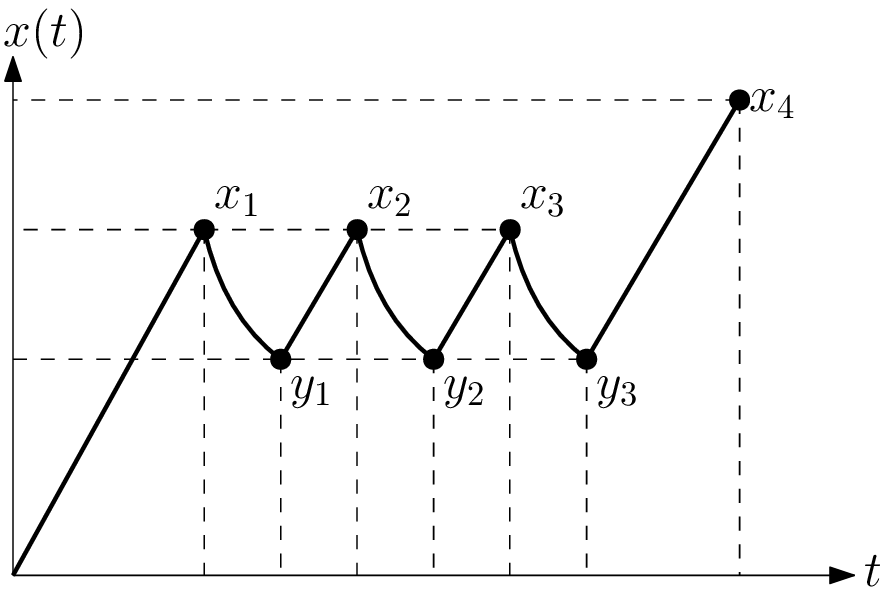}}
	\hfill
	\subfloat[\label{opt_soln_exp2}]{%
		\includegraphics[width=0.43\linewidth]{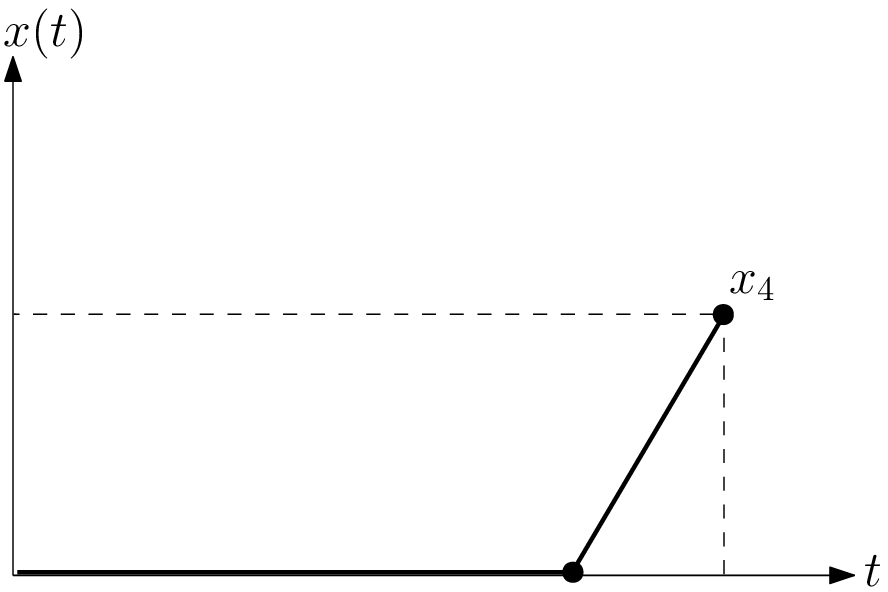}}
	\caption{ Optimal solution for the exponentially decaying age case: (a) When $T_c < T-\frac{1}{\alpha}$ (relatively small update duration). (b) When $T_c>T-\frac{1}{\alpha}$ (relatively large update duration).}
   \label{opt_soln_all_exp}
\end{figure}	

Finally, we summarize the optimal policy for the exponentially decaying age case combining the results in Sub-sections~\ref{exponential-x1-positive} and \ref{exponential-x1-zero}. If $T_c< T-\frac{1}{\alpha}$, i.e., the allowed update duration is relatively small with respect to the total session duration, then the optimal policy is to update $N$ times with equal update durations $c_i=\frac{T_c}{N}.$ Also, in this case, all $x_i$ for $i=1,\dots,N$ should be equal as given in (\ref{x1_closed}), and all $y_i$ for $i=1,\dots,N$ should be equal as well. An example age evolution curve for this case for $N=3$ is shown in Fig.~\ref{opt_soln_all_exp}(a). If $T_c>T-\frac{1}{\alpha}$, i.e., the allowed update duration is relatively large compared to the total session duration, then the optimal policy is to update starting from $t=0$ till $t=T_c$, and then let the age grow afterwards until $t=T$. There are multiple optimal assignments of total update duration $T_c$ to $c_i$ in this case; we choose $c_i = \frac{T_c}{N}$ again for symmetry with the previous case. Also, in this case, all $x_i$ for $i=1,\dots,N$ are equal and equal to zero, and all $y_i$ for $i=1,\dots,N$ are equal and equal to zero as well. An example age evolution curve for this case is shown in Fig.~\ref{opt_soln_all_exp}(b).

\section{Linearly Decaying Age Model} \label{sect:linear}

\begin{figure}[t]
	\centerline{\includegraphics[width=0.7\columnwidth]{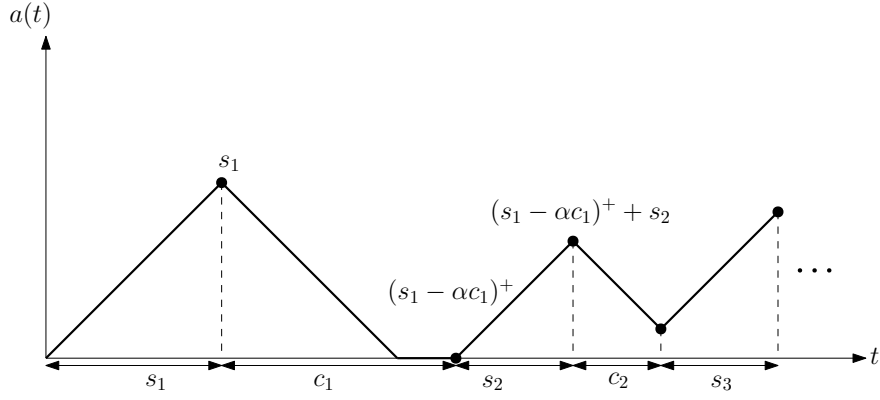}}
	\caption{A general example evolution of age in the case of linearly decaying age.}
	\label{Fig1}
\end{figure}

In this section, we consider the linearly decaying age model where the aging process can be slower or faster than the updating process. We consider the most general case by allowing the slope in the soft update policy, $\alpha$, to be arbitrary. In additional, when the duration of soft update process is sufficiently large, the instantaneous age can be reduced to zero. In this case, we can further continue the soft update process, and as a result, keep the age at the level of zero, i.e., not allow it to grow. A general example evolution of age for the linearly decaying age model is shown in Fig.~\ref{Fig1}. Age, in this case, is given as:
\begin{align}
A_T = &\frac{\alpha +1}{2\alpha}\sum_{i=1}^{N}\left(\left(s_i +\sum_{j=0}^{i-1}\left(s_j-\alpha c_j\right)^+\right)^2-\left(\sum_{j=1}^{i}\left(s_j-\alpha c_j\right)^+\right)^2\right)\nonumber\\
&+\frac{(s_{N+1}+\sum_{j=1}^{N}\left(s_j-\alpha c_j\right)^+)^2}{2} \label{AT-linear}
\end{align}
where $c_0=0$, $s_0=0$, and $s_{N+1}=T-\sum_{i =1}^{N}(s_i+c_i)$.

Next, we identify some important properties of the optimal solution. First, the following lemma states that, in the optimal solution, total update time, $T_c$, should be completely utilized.

\begin{lemma}\label{lemma_1}
	For the linearly decaying age model, in the optimal policy, we must have $\sum_{i=1}^{N} c_i = T_c$.
\end{lemma}

\begin{Proof}
	We prove this by contradiction. Assume that in the optimal policy, we have $\sum_{i=1}^{N} c_i < T_c$. First, let us choose the smallest index, $j$, such that $a(t_j')>0$. We can decrease the age further by increasing $c_j$. This policy is still feasible since the total update time constraint is not tight. Thus, we continue to increase $c_j$ until either $a(t_j') = 0$ or $\sum_{i=1}^{N} c_i = T_c$. If $a(t_j') = 0$ and $\sum_{i=1}^{N} c_i < T_c$, we move to the second smallest index such that the age at the end of the update period is not zero and apply the same procedure. We apply this procedure until $a(t_i') = 0$ for all $i$. At the end, if we obtain $\sum_{i=1}^{N} c_i < T_c$ and $a(t_i') = 0$ for all $i$, we can further decrease the age by increasing the duration of any update process by the amount $T_c-\sum_{i=1}^{N} c_i$. Since $a(t_i') = 0$ for all $i$, the age will stay at zero. Thus, we obtain a new policy where $\sum_{i=1}^{N} c_i =T_c$. This new policy has smaller age at each step, implying we have reached a contradiction. Thus, in the optimal policy, $\sum_{i=1}^{N} c_i = T_c$.	
\end{Proof}

From Lemma \ref{lemma_1}, we see that the total update time, $T_c$, should be fully used. Thus, when $T_c =T$, the optimal solution is to update the system starting from $t=0$ to $t=T$. The optimal age in this case is $A_T=0$. When $T_c<T$, we have time intervals where the system ages. If we decrease $T_c$, the total time where the age stays at zero decreases since there will be no update for $T-T_c$ and some portion of an update process can be used to decrease the age to zero. Let us first consider the case where $\sum_{i=1}^{k}\left(s_i-\alpha c_i\right) \geq 0$, for all $k=1,\dots,N$. In other words, we consider the case where each soft update process ends before or as soon as instantaneous age reaches zero. After providing a solution for this specific case, we generalize the solution to the most general case where the age can stay at zero. Thus, we formulate the problem with this condition enforced, as follows:
\begin{align}
\label{problem_2}
\min_{\{s_{i}, c_{i} \}}  \quad & A_T \nonumber \\
\mbox{s.t.} \quad & \sum_{i=1}^{N+1} s_i+\sum_{i=1}^{N} c_i = T \nonumber  \\
\quad & \sum_{i=1}^{N} c_i \leq T_c \nonumber  \\
\quad &\sum_{i =1}^{k} s_i-\alpha c_i\geq 0, \quad \forall k
\end{align}
where $A_T$ in the cost function is the age expression in (\ref{AT-linear}); the first constraint is the total session duration constraint which is the sum of aging and update durations; the second constraint is the constraint on the total soft update duration; and the third (last) constraint enforces that each update duration ends before or as soon as the age goes down to zero as discussed above.

This is not a convex optimization problem as the objective function is not convex. Our approach will be to lower bound the objective function, minimize this lower bound, and then show that this minimized lower bound can be achieved with a certain feasible selection of the variables. First, the following lemma states that, in the optimal solution, the age should be equal to zero at the end of each and every soft update period, i.e., the update period should never end before the age goes down exactly to zero.

\begin{lemma}\label{lemma_2}
	For the linearly decaying age model, for the problem in (\ref{problem_2}) which terminates updates if the age reaches zero, in the optimal policy, the age should be exactly equal to zero at the end of each soft update period i.e., $a(t_i') = 0$ for all $i$. In addition, $c_i = \frac{T_c}{N}$, $s_i =\frac{\alpha T_c}{N}$ for all $i=1,\dots,N$, and $s_{N+1} = T- (\alpha+1)T_c$.
\end{lemma}

\begin{Proof}
	 We first note that $A_T$ in (\ref{AT-linear}) can equivalently be written as:
	\begin{align}
	A_T =&\frac{\alpha+1}{2}\left(\alpha\sum_{i=1}^{N} c_i^2+2\sum_{i=1}^{N} (s_i-\alpha c_i)\left(\sum_{j=i}^{N} c_j\right)\right)+ \frac{(T-(\alpha+1)T_c)^2}{2} \label{AT-linear-equiv}
	\end{align}	
	We next note that, even though we do not know the sign of each $(s_i-\alpha c_i)$ in (\ref{AT-linear-equiv}) at this point, we know that the entirety of the middle term in (\ref{AT-linear-equiv}) is always non-negative since:
	\begin{align}
		\sum_{i=1}^{N}(s_i-\alpha c_i)\left(\sum_{j=i}^{N}c_j\right) = \sum_{i=1}^N \left(\sum_{j=1}^i s_j-\alpha c_j\right)c_i
	\end{align}
	where the right hand side is non-negative due to the constraints in (\ref{problem_2}). Thus, we lower bound (\ref{AT-linear-equiv}) by setting the middle term as zero by choosing $s_i=\alpha c_i$ for all $i$ which also implies that the age is equal to zero at the end of each soft update period.
	Then, minimizing the lower bound becomes equivalent to minimizing $\sum_{n=1}^{N} c_i^2$ subject to $\sum_{i=1}^{N}c_i = T_c$, whose solution is $c_i = \frac{T_c}{N}$. Then, we can choose $s_i=\alpha c_i$ and $c_i = \frac{T_c}{N}$ for all $i=1,\dots, N$, and $s_{N+1}= T-(\alpha +1)T_c$.
\end{Proof}

Next, we extend our solution to include the cases where the age can stay as zero. Towards that end, in the following lemma, we prove that the age cannot stay at zero for some update process(es) unless age becomes zero at the end of each and every update.

\begin{lemma}\label{lemma_2_add}
	For the linearly decaying age model, in the optimal policy, if the age stays at zero for some update process(es), then the age should be equal to zero after each update period.
\end{lemma}

\begin{Proof}
We prove this by contradiction. Assume that we have an optimal update policy where the age stays at zero for a total of $T_0$ amount of time and yet there exists an update period $i$ where $s_i-\alpha c_i>0$, i.e., the age does not go down to zero after the $i$th update period. Then, subtract $T_0$ from the total update duration $T_c$, and consider the age minimization problem with a total update duration of $T_c'=T_c-T_0$. We know from Lemma~\ref{lemma_2} that if the age does not decrease down to zero after each update, the update policy cannot be optimal. Therefore, there exists a policy which yields a smaller age than the assumed optimal update policy. Thus, we have reached a contradiction and the original update policy cannot be optimal. Hence, if the age stays at zero for some update process(es), then the age should be equal to zero after each update.
\end{Proof}

Next, we find the optimal solution structure for the case where the age stays at zero for some update process(es).

\begin{lemma}\label{lemma_3}
	For the linearly decaying age model, in the optimal policy, if the age stays at zero for some update process(es), then the optimal policy is to choose $c_i=\frac{T_c}{N}$ and $s_{i}=\frac{(T-T_c)\alpha}{\alpha (N+1)+1}$ for $i=1,\dots,N$, and $s_{N+1}=\frac{(T-T_c)(\alpha+1)}{\alpha(N+1)+1}$. In addition, we must have $T_c\geq \frac{NT}{(\alpha+1)(N+1)}$.
\end{lemma}

\begin{Proof}
Since we consider the case where the age stays at zero, age at the end of each update process should be equal to zero due to Lemma~\ref{lemma_2_add}. Thus, $A_T$ in (\ref{AT-linear}) becomes:
\begin{align}
A_T = \frac{\alpha+1}{2\alpha}\sum_{i=1}^{N}s_i^2+\frac{s_{N+1}^2}{2}
\end{align} 	
For this case, we need to solve the following problem:
\begin{align}
\label{problem_3}
\min_{\{s_{i}, c_{i} \}}  \quad & \frac{\alpha+1}{2\alpha}\sum_{i=1}^{N}s_i^2+\frac{s_{N+1}^2}{2} \nonumber \\
\mbox{s.t.} \quad & \sum_{i=1}^{N+1} s_i= T-T_c\nonumber  \\
\quad & s_i-\alpha c_i\leq 0, \quad \forall i
\end{align}
The last constraint in (\ref{problem_3}) makes sure that age goes down to zero after each soft update period. We solve this problem using a Lagrangian:
\begin{align}
\mathcal{L} = \frac{\alpha+1}{2\alpha} \sum_{i=1}^{N}s_i^2+\frac{s_{N+1}^2}{2}-\lambda\left( \sum_{i=1}^{N+1} s_i  -T+T_c\right)
\end{align}
Taking the derivative with respect to $s_i$ and equating to zero, we obtain $s_i=\frac{\alpha\lambda}{\alpha+1}$ for $i=1,\dots,N$, and $s_{N+1}=\lambda$. Since $\sum_{i=1}^{N+1} s_i = T-T_c$, the optimal solution is $s_i = \frac{(T-T_c)\alpha}{\alpha (N+1)+1}$ for $i=1,\dots,N$, and $s_{N+1}=\frac{(T-T_c)(\alpha+1)}{\alpha(N+1)+1}$. Due to the last constraint, we must have $s_i= \frac{(T-T_c)\alpha}{\alpha (N+1)+1} \leq \alpha c_i$. Even though these constraints are satisfied by multiple sets of $c_i$'s, we choose the one with $c_i=\frac{T_c}{N}$. Finally, we need $T_c\geq \frac{NT}{(\alpha+1)(N+1)}$ in order to have feasible selections of $s_i \leq \alpha c_i$ for all $i$.
\end{Proof}

\begin{figure}
	\centering
	\subfloat[\label{Figure1}]{%
		\includegraphics[width=0.48\linewidth]{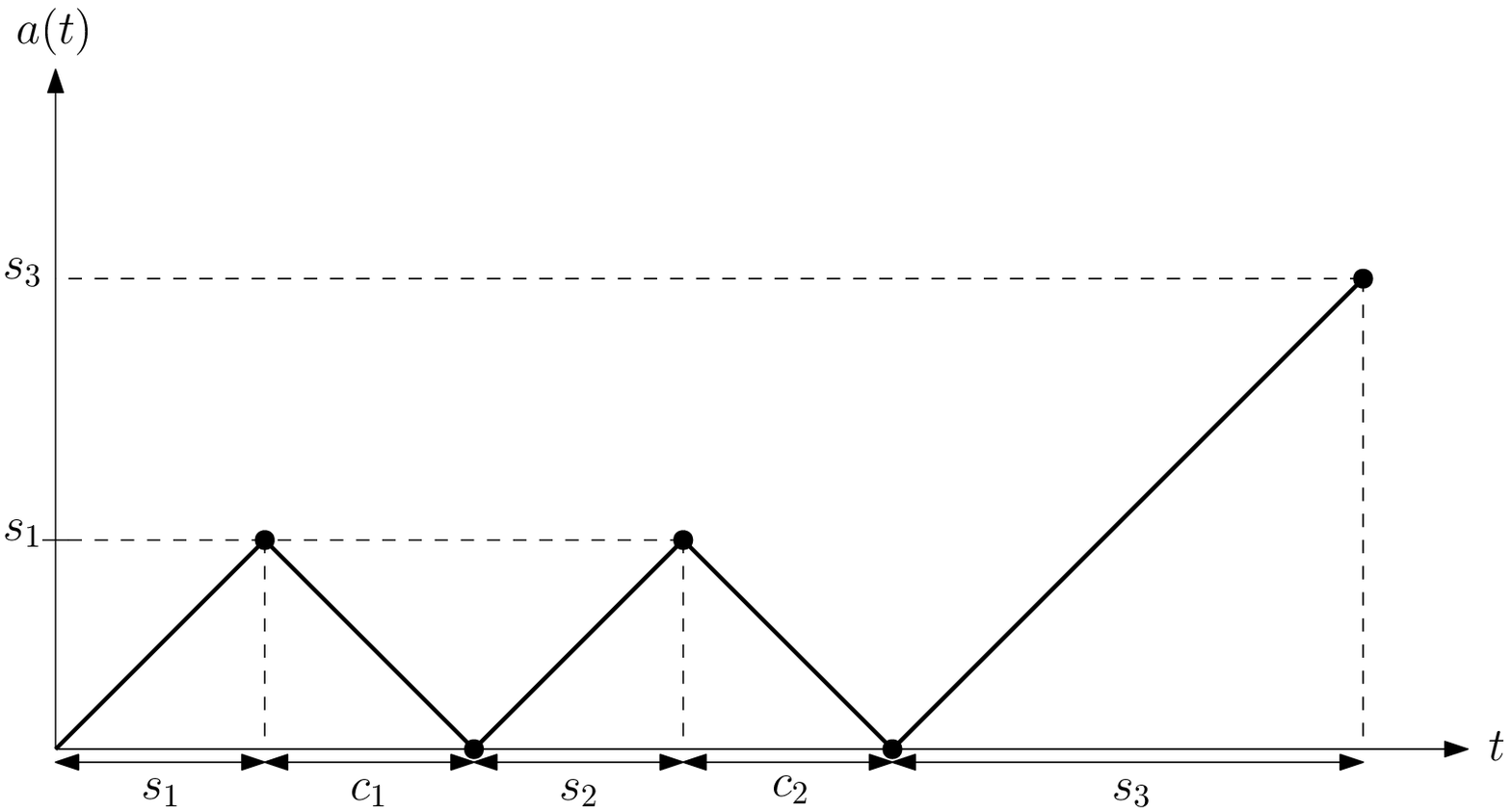}}
	\hfill
	\subfloat[\label{Figure3}]{%
		\includegraphics[width=0.48\linewidth]{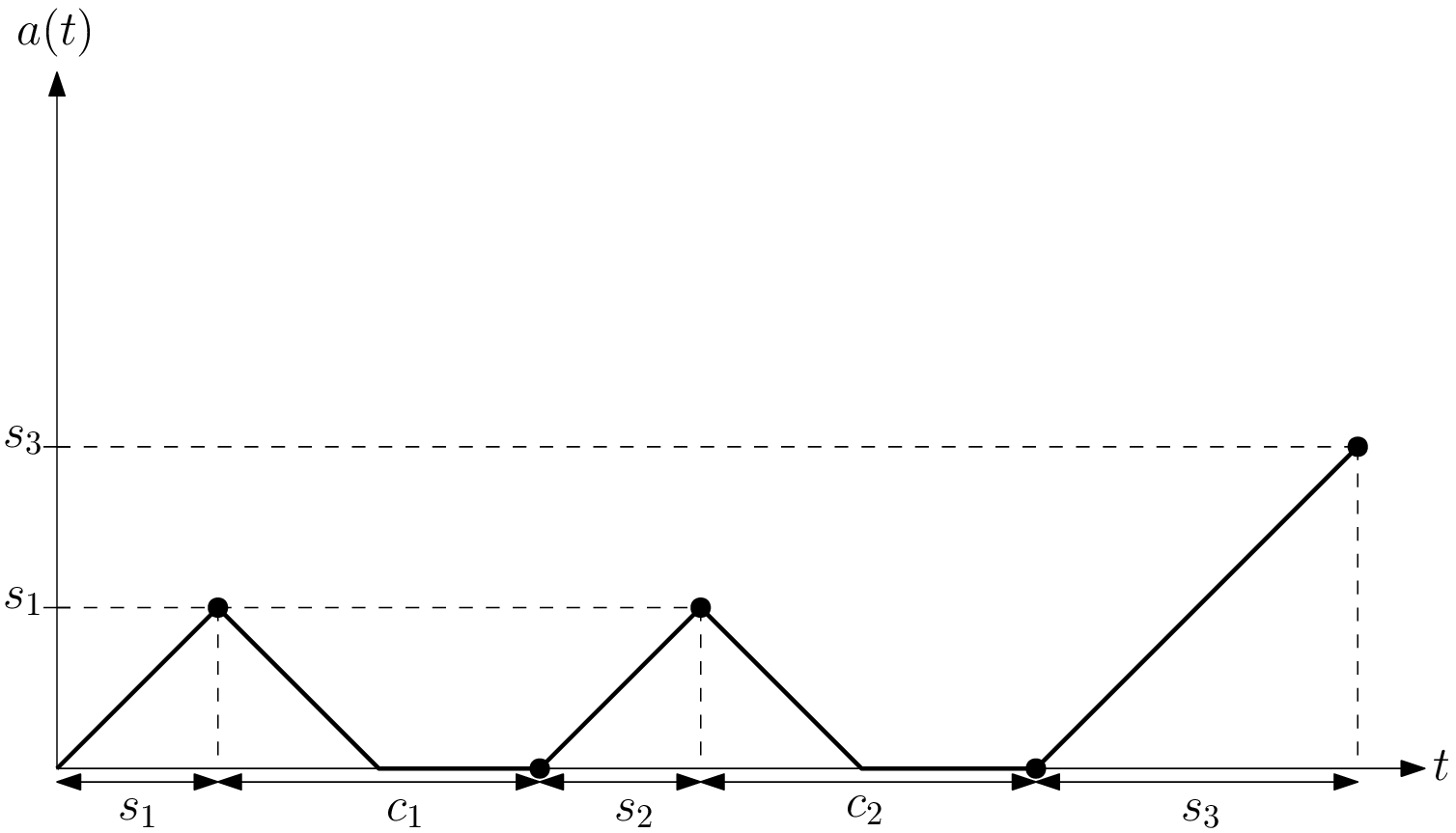}}
	\caption{Optimal policy structure for the linearly decaying age case: (a) When $T_c < \frac{NT}{(\alpha+1)(N+1)}$ and $\alpha =1$. (b) When $T_c \geq \frac{NT}{(\alpha+1)(N+1)}$ and $\alpha =1$. }
	\label{Figure1and3}
\end{figure}

Finally, we summarize the optimal policy for the linearly decaying age case. If $T_c < \frac{NT}{(\alpha+1)(N+1)}$, i.e., the allowed update duration is relatively small with respect to the total session duration, we are in Lemma~\ref{lemma_2} and the optimal policy is to choose $s_i=\alpha c_i$ and $c_i = \frac{T_c}{N}$ for $i = 1,\dots, N$, and $s_{N+1} = T-(\alpha+1)T_c $. An example age evolution curve for this case for $N = 2$ is shown in Fig.~\ref{Figure1and3}(a). If $T_c \geq \frac{NT}{(\alpha+1)(N+1)}$, i.e., the allowed update duration is relatively large compared to the total session duration, we are in Lemma~\ref{lemma_3} and the optimal policy is to choose $s_i = \frac{(T-T_c)\alpha}{\alpha (N+1)+1}$, $c_i = \frac{T_c}{N}$ for $i=1,\dots,N$, and $s_{N+1}=\frac{(T-T_c)(\alpha+1)}{\alpha(N+1)+1}$.\footnote{In \cite[Section~IV.B]{soft_upt_allerton}, the same result for $\alpha=1$ should hold. Therefore, when $T_c < \frac{NT}{2N+2}$, the solution remains the same as in \cite[Lemma~3]{soft_upt_allerton}. When $T_c \geq \frac{NT}{2N+2}$, the optimal solution is to choose $c_i=\frac{T_c}{N}$ and $s_{i}=\frac{T-T_c}{N+2}$ for $i=1,\dots,N$, and $s_{N+1}=\frac{2(T-T_c)}{N+2}$.} An example age evolution curve for this case for $N = 2$ is shown in Fig.~\ref{Figure1and3}(b). The optimal policy is to update exactly $N$ times in both cases with the age going down exactly to zero after each update. In addition, if the total update duration $T_c$ is large compared to the total time $T$ then the age stays at zero for some time for all update periods. Finally, we note that the case of age not going down to zero after the second update in the example general age evolution curve shown in Fig.~\ref{Fig1} can never happen.

\begin{figure}[t]
	\centerline{\includegraphics[width=0.6\columnwidth]{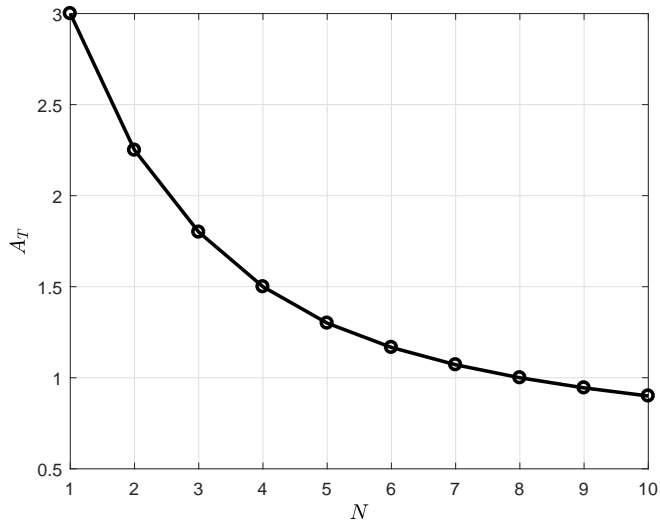}}
	\caption{Minimum age as a function of $N$ in the linearly decaying age case for $T=5$, $T_c =2$ $\alpha=1$.}
	\label{Sim4}
\end{figure}

Next, we investigate how the final minimum age expression varies as a function of the number of soft update opportunities $N$. If $T_c < \frac{NT}{(\alpha+1)(N+1)}$, the minimum age is:
\begin{align}
A_T= \frac{T_c^2}{N} \frac{\alpha(\alpha+1)}{2}+\frac{(T-(\alpha+1)T_c)^2}{2} \label{min-age1}
\end{align}
and if $T_c \geq \frac{NT}{(\alpha+1)(N+1)}$, the minimum age is:
\begin{align}
A_T= \frac{(\alpha+1)(T-T_c)^2}{2(\alpha (N+1)+1)}
\end{align}
For both cases, we observe that $A_T$ is a decreasing function of $N$. As an example, the minimum age as a function of $N$ is plotted in Fig.~\ref{Sim4} for $T=5$, $T_c=2$, $\alpha =1$.

\begin{figure}[t]
	\centerline{\includegraphics[width=0.6\columnwidth]{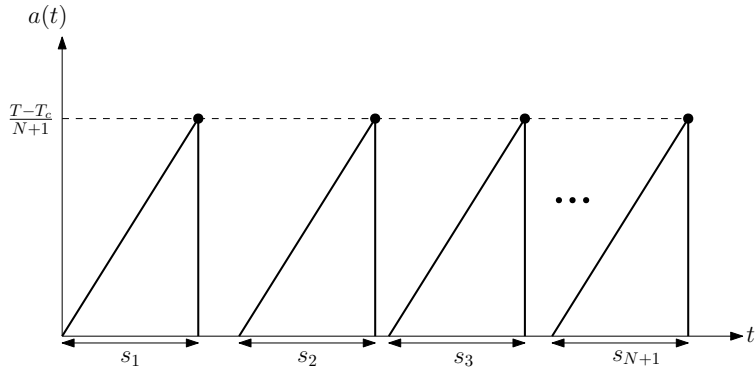}}
	\caption{Evolution of the optimal age when $\alpha\to\infty$.}
	\label{Fig3}
\end{figure}

Finally, we note that, when $\alpha\rightarrow \infty$, $T_c$ is only used to keep the age $a(t)= 0$, and the optimal age can be calculated as:
\begin{align}
\lim_{\alpha\to\infty} A_T = \frac{1}{2} \left(\frac{T-T_c}{N+1}\right)^2(N+1)
\end{align}
In this case, the optimal age is as shown in Fig.~\ref{Fig3}, which corresponds to the optimal age with instantaneous drops as in the existing literature except for the time intervals where the age stays at zero.\footnote{We observe that when $\alpha\to\infty$, the heights of the triangles become the same, which is similar to the result in \cite{soft_upt_allerton}.}

\section{Numerical Results} \label{sect:numresult}
\begin{figure}
	\centering
	\subfloat[\label{T_5_Tc_3_N_2}]{%
		\includegraphics[width=0.45\linewidth]{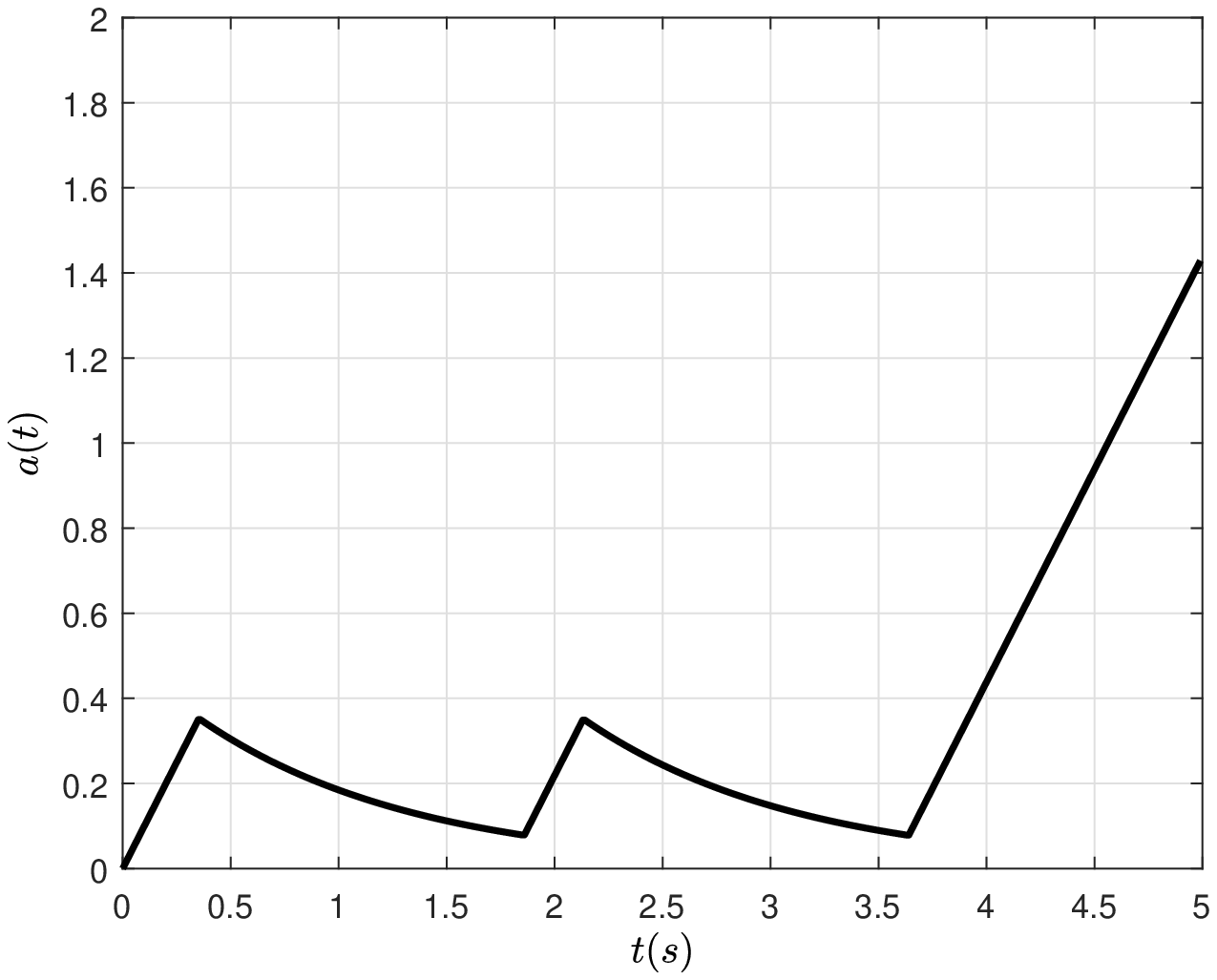}}
	\subfloat[\label{T_6_Tc_5_N_2}]{%
		\includegraphics[width=0.45\linewidth]{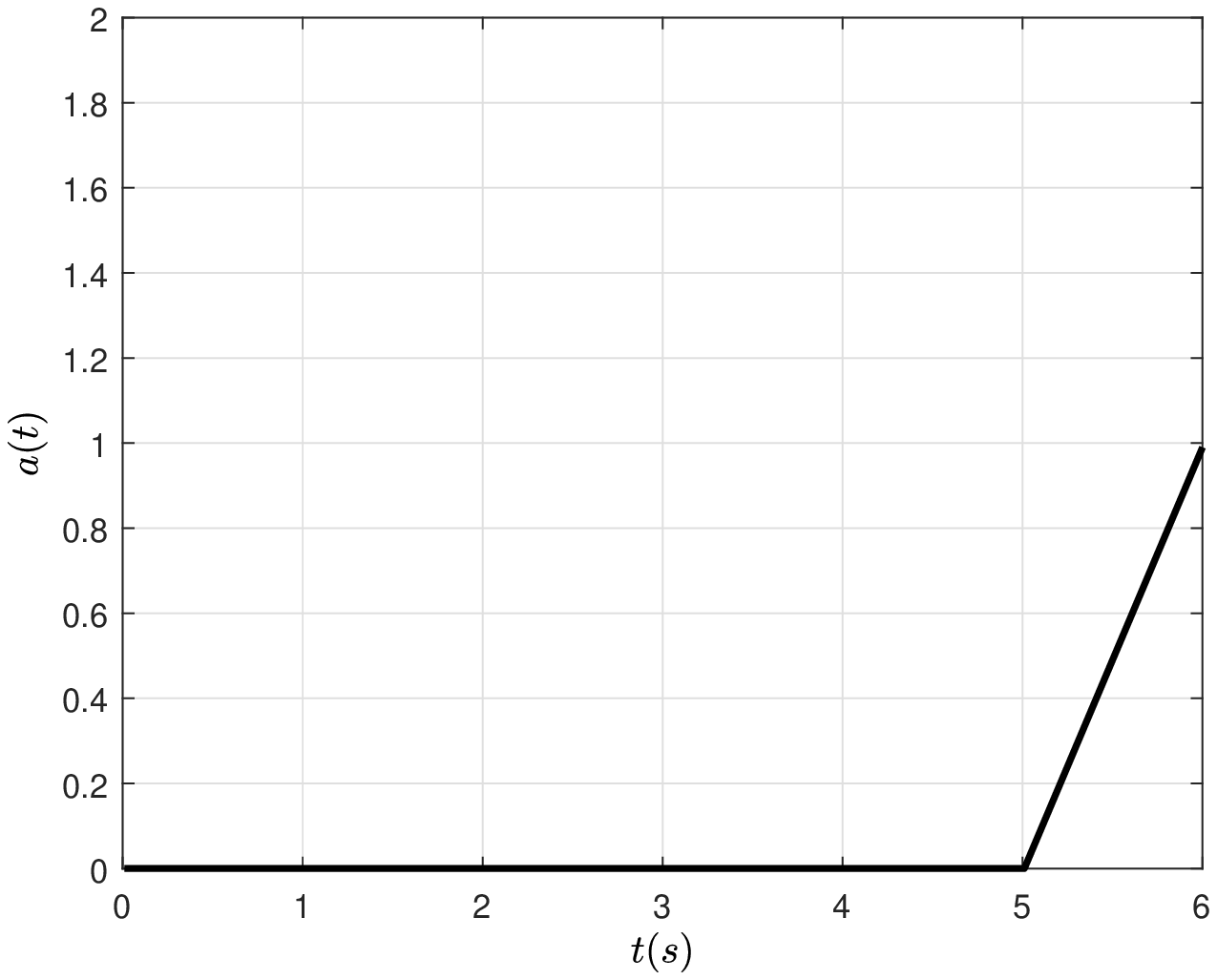}}\\
	\caption{Evolution of $a(t)$ in the exponentially decaying age model (a) when $N=2$, $T=5$, $T_c =3$, and $\alpha=1$,  (b) when $N=2$, $T=6$, $T_c =5$, and $\alpha=1$.}
\label{sim-exp}
\end{figure}

In this section, we give simple numerical examples to illustrate our results. In the first example, we consider the exponentially decaying age model with $T=5$, $T_c =3$, $N=2$ and $\alpha =1$. Since $T > T_c-\frac{1}{\alpha}$, the optimal update policy is to update $N=2$ times with equal time allocated to each update, i.e., $c_1 =c_2 = 1.5$. The evolution of $a(t)$ is shown in Fig.~\ref{sim-exp}(a).

In the second example, we consider the exponentially decaying age model with $T=6$, $T_c =5$, $N=2$ and $\alpha = 1$.
Since $T_c$ is large enough, i.e., $T\leq T_c-\frac{1}{\alpha}$, the system starts updating at $t=0$, proceeds to update continuously until $T_c$, and lets age grow then on until the end. The evolution of $a(t)$ is shown in Fig.~\ref{sim-exp}(b).

In the following three examples (third, fourth and fifth), we consider the linearly decaying age model with $\alpha = 1$. In the third example, we see the case where $T_c = \frac{NT}{N(\alpha+1)+\alpha}$. Note that if we have additional updating time, there will be time intervals where the age will stay at zero. The evolution of $a(t)$ is shown in Fig.~\ref{sim-lin}(a).

In the fourth example, we consider the case in Lemma~\ref{lemma_3}, where $T_c > \frac{NT}{(\alpha+1)(N+1)}$. We see that since $T_c$ is large enough compared to $T$, some of the total update time is used to make the age zero and for the remaining part of $T_c$, age will stay at zero which is shown in Fig.~\ref{sim-lin}(b).

In the fifth example, we consider the case where $T_c < \frac{NT}{(\alpha+1)(N+1)}$. In this case, age at the end of each update period is equal to zero. Since $T_c$ is small compared to $T$, in the optimal policy, we do not see any time intervals where the age stays at zero. The evolution of $a(t)$ is shown in Fig.~\ref{sim-lin}(c).

So far, we have provided examples for the linear case with $\alpha = 1$. In the following examples, we consider the cases with $\alpha>1$ and $\alpha<1$. In the first case, we choose $\alpha = 2$, $N=2$, $T=3$, and $T_c =0.8$, and in the second case, we choose $\alpha = 0.5$, $N=2$, $T=3.6$, and $T_c =1.6$. The optimal policies are shown in Fig.~\ref{sim-lin-alpha}(a) and Fig.~\ref{sim-lin-alpha}(b), respectively.

\section{Conclusion and Future Directions} \label{sect:rest_upt}
In this paper, we introduced the concept of soft updates which is relevant in systems with human interactions and social media settings, where the decrease in age happens gradually over soft update periods. We study two soft update regimes: in the first one, age decays exponentially and in the second one age decays linearly during the soft update period. In both models, we showed that the optimal policy is to have $N$ updates and $T_c$ should be completely utilized with allocating equal amount of time for each update.

\begin{figure}
	\centering
	\subfloat[\label{Sim1}]{%
		\includegraphics[width=0.45\linewidth]{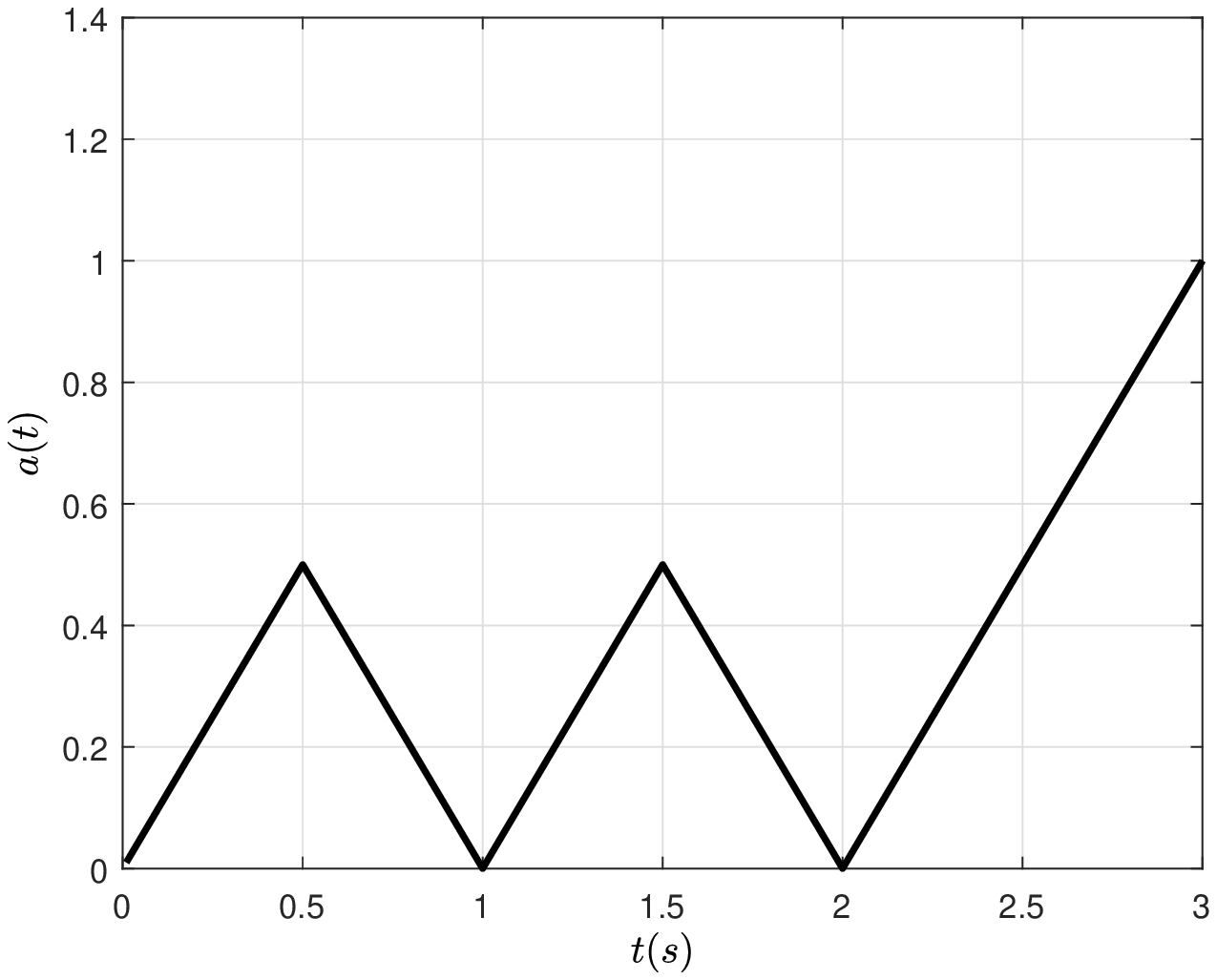}}\\
	\subfloat[\label{Sim2}]{%
		\includegraphics[width=0.45\linewidth]{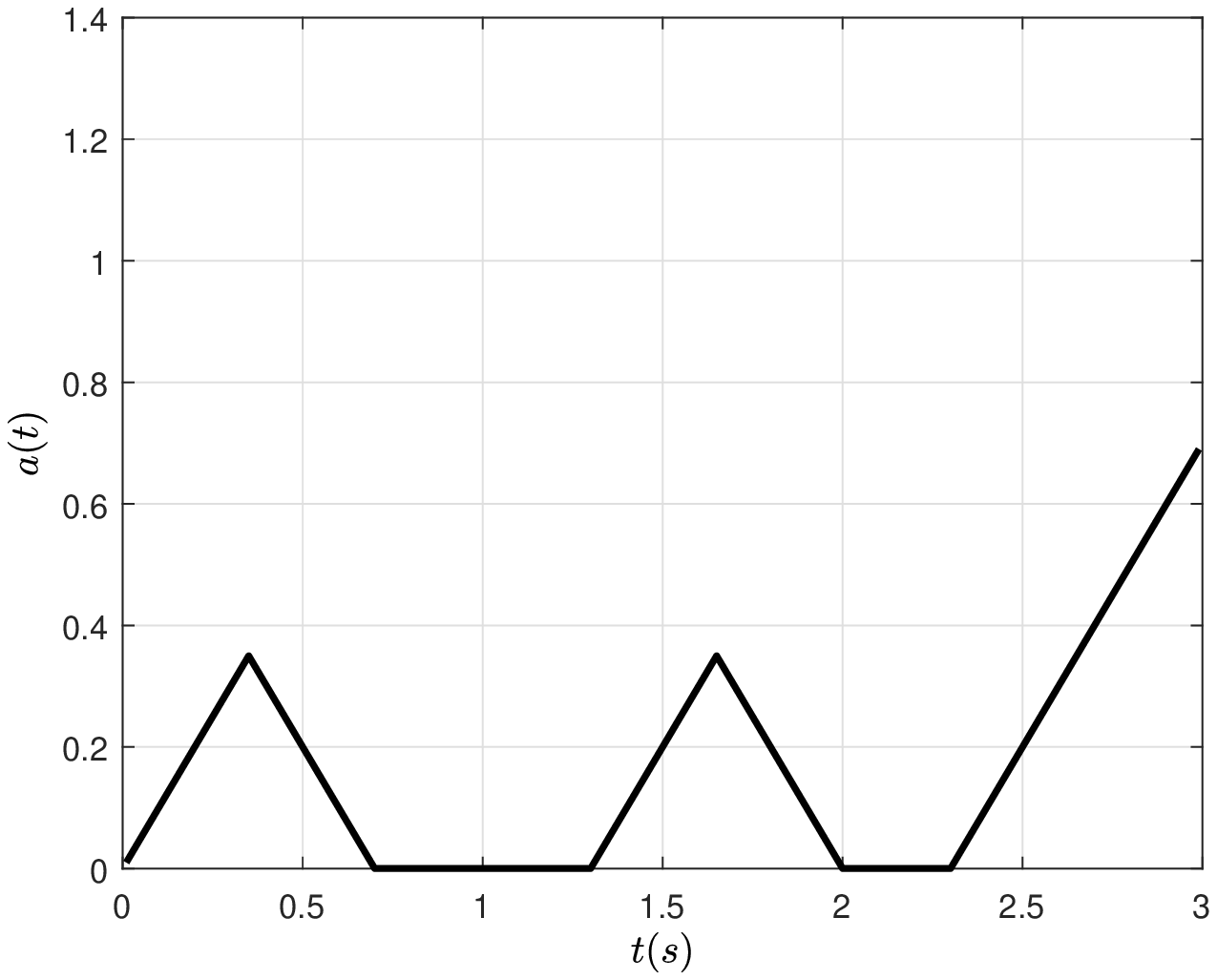}}
	\hfill
	\subfloat[\label{Sim3}]{%
		\includegraphics[width=0.45\linewidth]{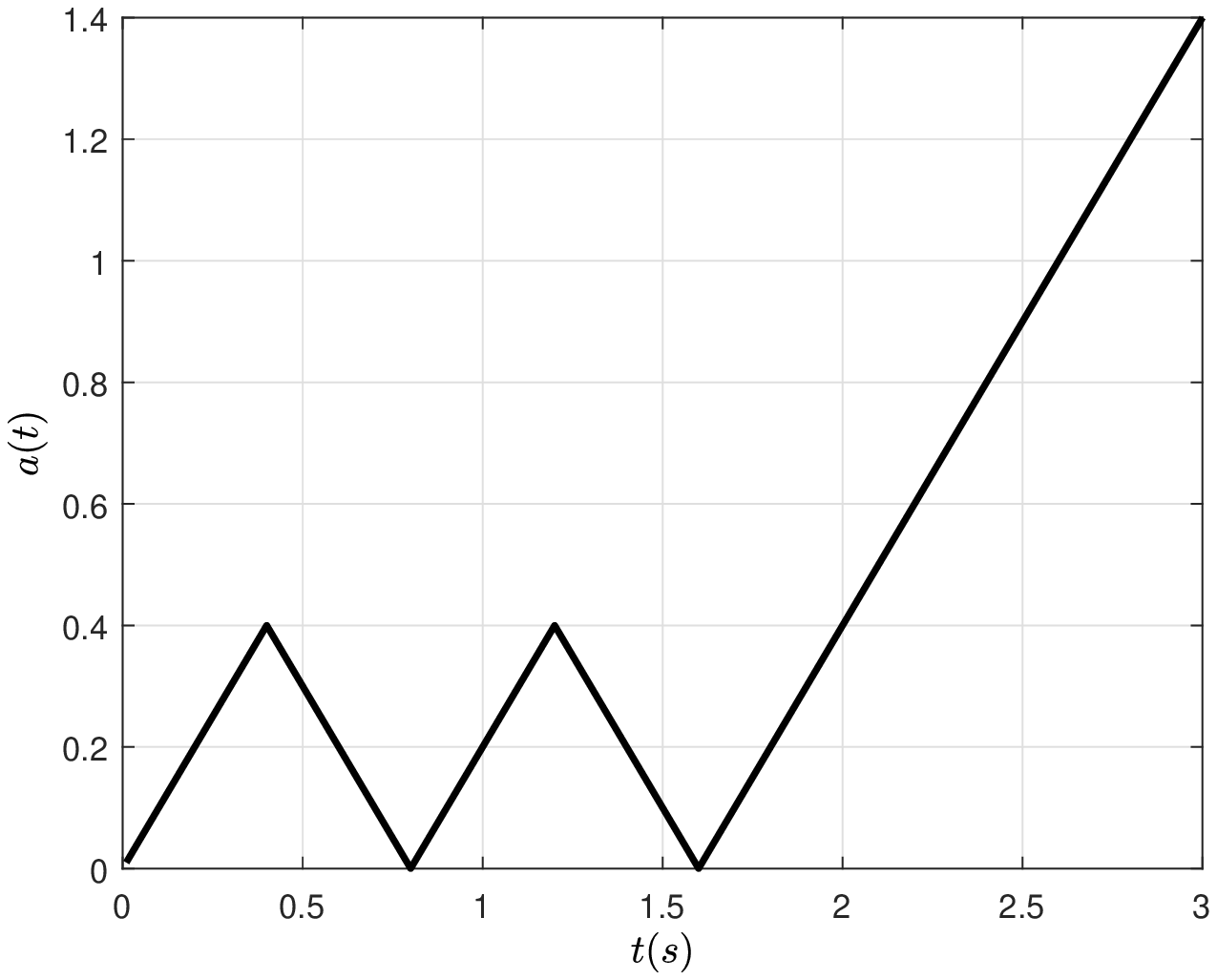}}
	\caption{Evolution of $a(t)$ in the linearly decaying age model, for $\alpha=1$, $N=2$, $T=3$, and (a) $T_c =1$, (b) $T_c =1.6$, (c) $T_c =0.8$.}
\label{sim-lin}
\end{figure}

\begin{figure}
	\centering
	\subfloat[\label{T_3_Tc_0_8_alpha_2}]{%
		\includegraphics[width=0.43\linewidth]{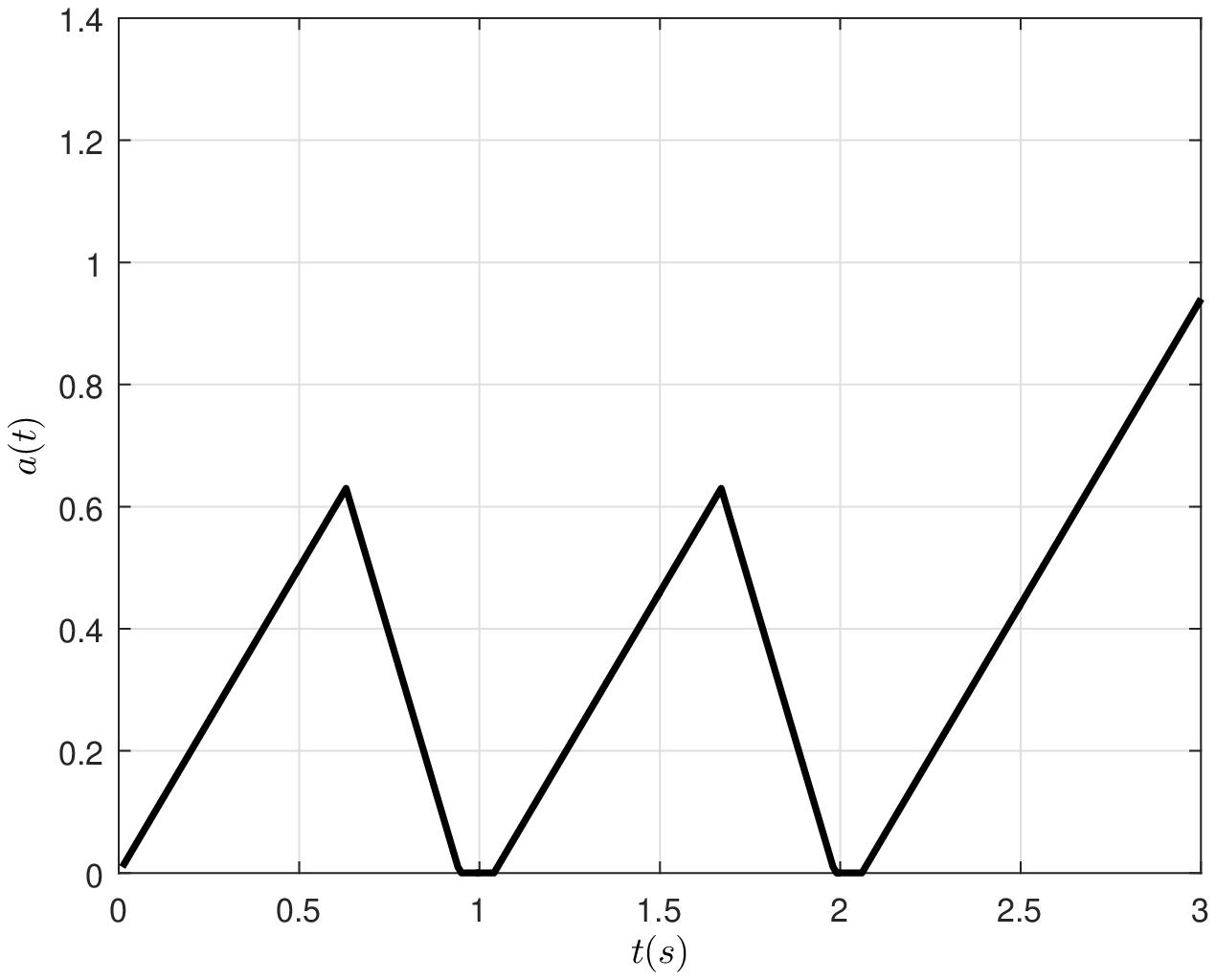}}
	\subfloat[\label{T_3_6_Tc_1_6_alpha_0_5}]{%
		\includegraphics[width=0.43\linewidth]{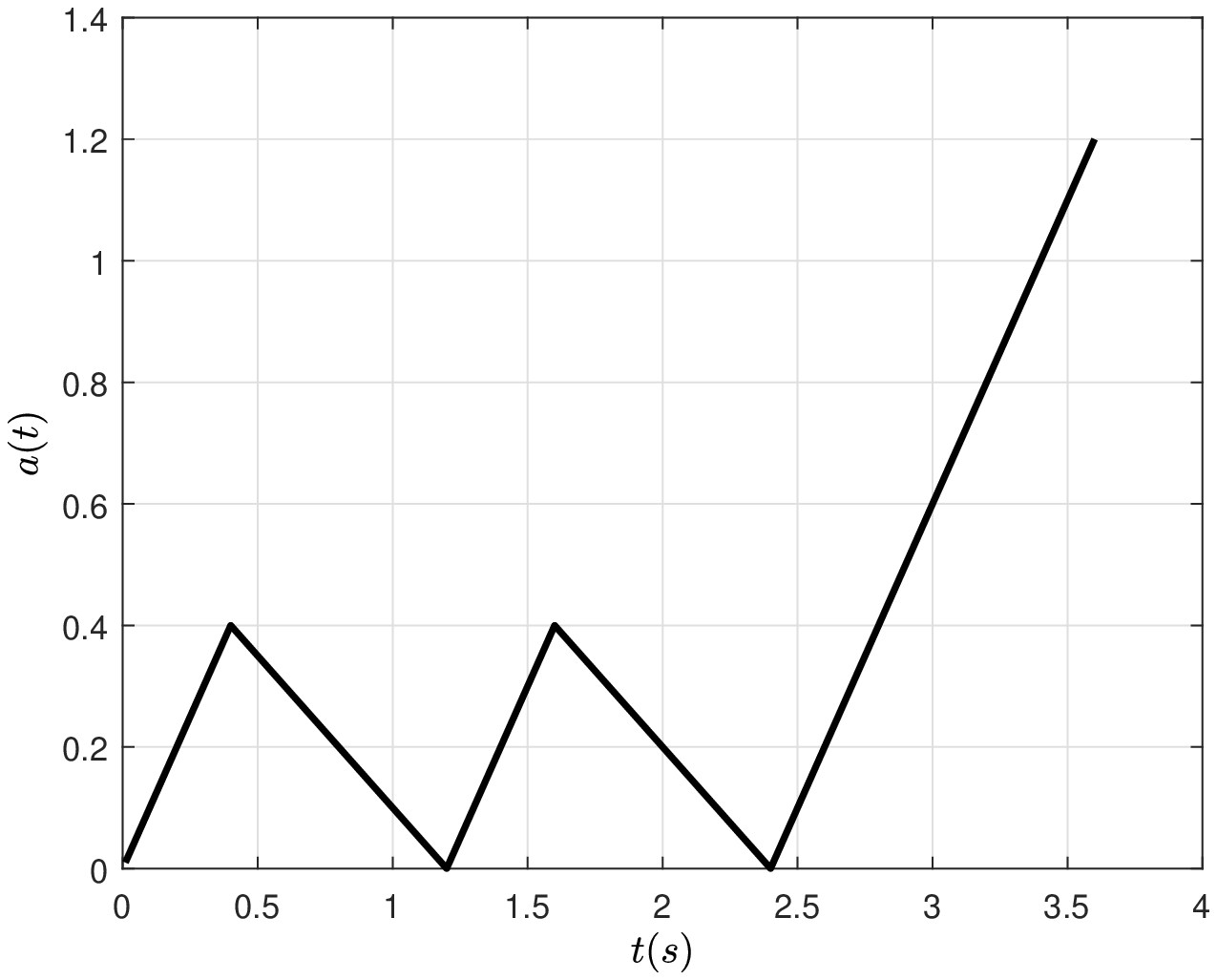}}\\
	\caption{Evolution of $a(t)$ in the linearly decaying age model (a) $\alpha =2$, $N=2$, $T=3$, and $T_c =0.8$, and (b) $\alpha =0.5$, $N=2$, $T=3.6$, and $T_c =1.6$.}
	\label{sim-lin-alpha}
\end{figure}
\begin{figure}
	\centering
	\subfloat[\label{Rest1}]{%
		\includegraphics[width=0.43\linewidth]{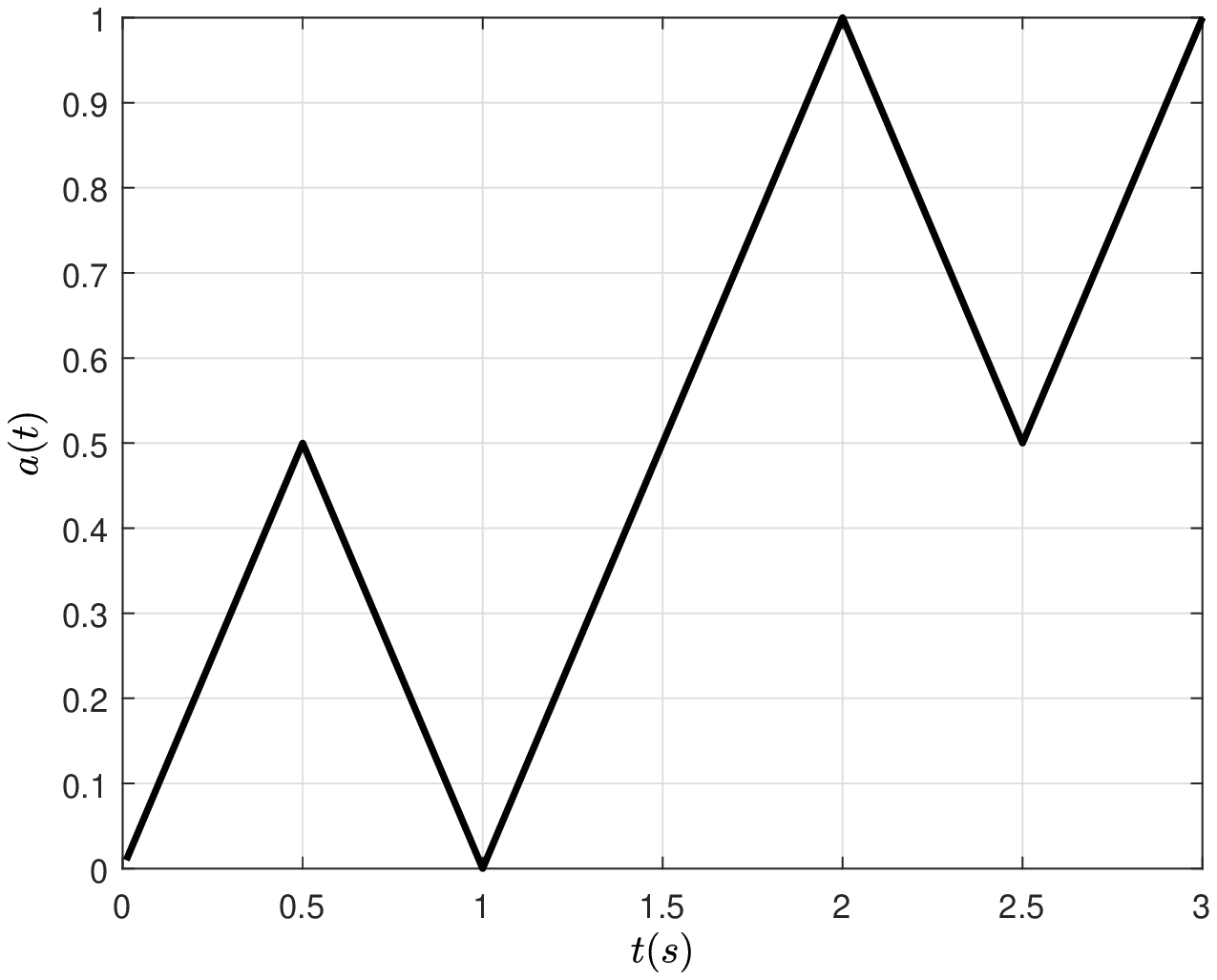}}
	\subfloat[\label{Rest2}]{%
		\includegraphics[width=0.43\linewidth]{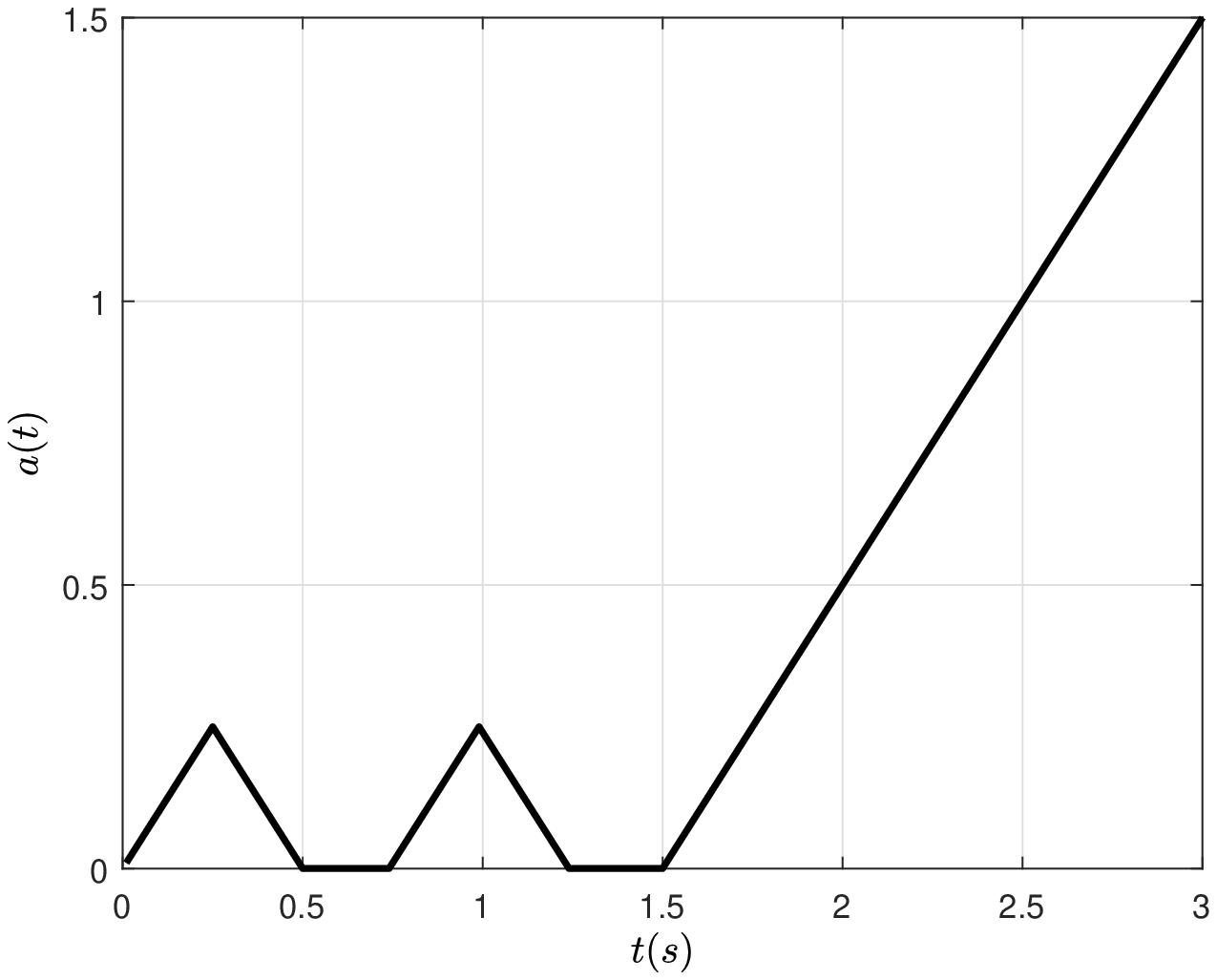}}\\
	\caption{Evolution of $a(t)$ in the linearly decaying age model, for $N=2$, $T=3$, $T_c =1$, and $\alpha = 1$. Updates are allowed in between (a) $t\in[0,1]$ and $t\in [2,3]$, and (b) $t\in[0,1.5]$ and $t\in [2,3]$. }
	\label{sim-future}
\end{figure}
For future work, further restrictions on the update times can be considered, e.g., restrictions on the time intervals in which meetings may take place. We provide two numerical results for these cases. In the first example, we consider the case where $T =3$, $T_c =1$, $N=2$, $\alpha =1$, and we restrict updates to take place only in the intervals $t\in[0,1]$ and $t\in [2,3]$. We recall that if there is no further restriction on the update processes, the optimal age evolution is given in Fig.~\ref{sim-lin}(a). Note that since updating is not allowed in between $t\in( 1,2)$, the optimal age evolution is different and is as shown in Fig.~\ref{sim-future}(a). In this case, we see that $T_c$ is fully used and age becomes zero at the end of the first update period which seems to follow the optimal policy structure with no restrictions. On the other hand, age is not equal to zero after the second update unlike the unrestricted case studied in this paper; see again Fig.~\ref{sim-lin}(a).

In the second example, we consider the same case except this time, updating is not allowed in between $t\in(1.5,2)$. In Fig.~\ref{sim-future}(b), we see the optimal age evolution in this case. Even though the updates are allowed in $t\in[2,3]$, the system chooses to use them at the beginning and the age becomes zero after each update. Note that in both examples $T_c$ is fully utilized but, in the first case, even though $T_c$ can be utilized at the beginning, i.e., $t\in[0,1]$ one of the updates takes place in between $t\in[2,3]$. It seems that there is a point at which keeping the maximum age smaller is more important whereas after this point, reducing the age to zero after each update yields an optimal solution.

\bibliographystyle{unsrt}
\bibliography{myLibrary}

\end{document}